%% file: bare_jrnl.tex
\documentclass[journal]{IEEEtran}
\usepackage{amsmath,amsfonts}
\usepackage{algorithmic}
\usepackage{array}
\usepackage[caption=false,font=normalsize,labelfont=sf,textfont=sf]{subfig}
\usepackage{textcomp}
\usepackage{stfloats}
\usepackage{url}
\usepackage{bm}
\usepackage{verbatim}
\usepackage{graphicx}
\usepackage{comment}
\usepackage{booktabs}
\usepackage{multirow}
\usepackage{mathrsfs}
\usepackage{color}
\usepackage{ulem}
\usepackage{flushend}
\newcommand{\add}[1]{\textcolor{black}{#1}}
\newcommand{\revise}[1]{\textcolor{black}{#1}}
\newcommand{\secondRevise}[1]{\textcolor{black}{#1}}
\newcommand{\remove}[1]{}

\def\BibTeX{{\rm B\kern-.05em{\sc i\kern-.025em b}\kern-.08em
    T\kern-.1667em\lower.7ex\hbox{E}\kern-.125emX}}
\usepackage{balance}

\begin{document}

\title{Design, Control, and \add{Motion-Planning} for \add{a} Root-\add{Perching} Rotor-\add{Distributed} Manipulator}
\author{Takuzumi Nishio$^{1}$, \rm{Moju Zhao}$^{2}$, \rm{Kei Okada}$^{1}$, \rm{Masayuki Inaba}$^{1}$
\thanks{This paper was recommended for publication by Editor Paolo Robuffo Giordano upon evaluation of the Associate Editor and Reviewers' comments. This work was supported by JSPS KAKENHI Grant ID 22710191.}
\thanks{$^{1}$ Department of Mechano-Infomatics, The University of Tokyo, $^{2}$ Department of Mechanical Engineering, The University of Tokyo, 7-3-1 Hongo, Bunkyo-ku, Tokyo 113-8656, Japan. ({\it corresponding author: Takuzumi Nishio},{\tt\small nishio-takuzumi@ynl.t.u-tokyo.ac.jp}).}
 \thanks{Digital Object Identifier (DOI): see top of this page.}}

\markboth{IEEE Transactions on Robotics,~Vol.~40, No.~9, September~2023}%
{How to Use the IEEEtran \LaTeX \ Templates}

\maketitle

\begin{abstract}
Manipulation performance improvement is crucial for aerial robots. For aerial manipulators, the \revise{baselink} position and attitude errors directly affect the precision at \revise{the} end effector. To address this stability problem, fixed-body approaches such as perching on the environment using the rotor suction force are useful. Additionally, conventional arm-equipped multirotors, called rotor-concentrated manipulator\revise{s} (RCM\revise{s}), find it difficult to generate a large wrench at the end effector due to joint torque limitations. 
Using distributed rotors to each link, the thrust can support each link weight, decreasing the arm joints' torque.
Based on this approach, rotor-distributed manipulators (RDMs) can increase feasible wrench and reachability \revise{of the end-effector}. 
This paper introduces a minimal configuration of a rotor-distributed manipulator that can perch on surfaces, especially ceilings, using a part of their body. First, we design a minimal rotor-distributed arm considering the flight and end-effector performance. Second, a flight controller is proposed for this minimal RDM along with a perching controller adaptable for various types of aerial robots. Third, we propose a motion planning method based on inverse kinematics (IK), considering specific constraints to the proposed RDMs such as \revise{perching force}. Finally, we evaluate flight and perching motions and \revise{confirm} that the proposed manipulator can significantly improve the manipulation performance.
\end{abstract}


\input 01-intro.tex

\input 02-design.tex
\input 03-model.tex
\input 04-control.tex
\input 05-planning.tex
\input 06-system.tex
\input 07-experiment.tex
\input 08-conclusion.tex

\input{appendix}
\bibliographystyle{unsrt}
\bibliography{reference}

\begin{IEEEbiography}[{\includegraphics[width=1in,height=1.25in,clip,keepaspectratio]{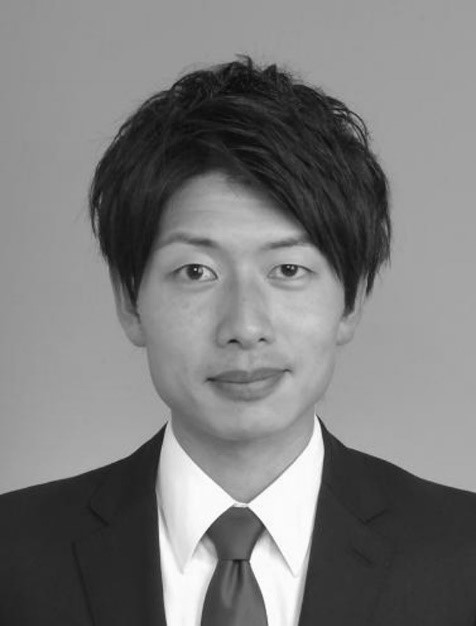}}]{Takuzumi Nishio}
 is an Assistant Professor in the Department of Mechano-Informatics, School
of Information Science and Technology, University of Tokyo. He received PhD degree from the Department of Interdisciplinary-Informatics, University of Tokyo, in 2022. His research interests are mechanical design, modeling and control of aerial robots, and vision-based recognition and motion planning of manipulators in field robotics.\end{IEEEbiography}

\begin{IEEEbiography}[{\includegraphics[width=1in,height=1.25in,clip,keepaspectratio]{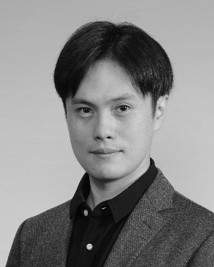}}]{Moju Zhao}
 is currently an Assistant Professor at The University of Tokyo. He received Doctor Degree from the Department of Mechano-Informatics, The University of Tokyo, 2018. His research interests are mechanical design, modelling and control, motion planning, and vision based recognition in aerial robotics. His main achievement is the articulated aerial robots which have received several awards in conference and journal, including the Best Paper Award in IEEE ICRA 2018.\end{IEEEbiography}

\begin{IEEEbiography}[{\includegraphics[width=1 in,height=1.25 in,clip,keepaspectratio]{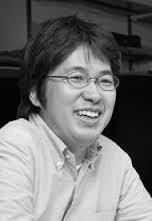}}]{Kei Okada}
 received BE in Computer Science from Kyoto University in 1997.
He received MS and PhD in Information Engineering from The University of Tokyo
in 1999 and 2002, respectively. From 2002 to 2006, he joined the Professional
Programme for Strategic Software Project in The University Tokyo. He was appointed
as a lecturer in the Creative Informatics at the University of Tokyo in 2006 and
an associate professor in the Department of Mechano-Informatics in 2009. His research interests include humanoids robots, real-time 3D computer vision, and recognitionaction integrated system.
\end{IEEEbiography}

 \begin{IEEEbiography}[{\includegraphics[width=1in,height=1.25in,clip,keepaspectratio]{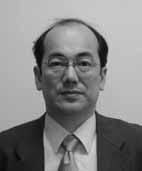}}]{Masayuki Inaba}
  is a professor in the Department of Mechano-Informatics, School
of Information Science and Technology, University of Tokyo. He graduated from the
Department of Mechanical Engineering, University of Tokyo, in 1981, and received
MS and PhD degrees from the Graduate School of Information Engineering, University of Tokyo, in 1983 and 1986. He was appointed as a lecturer in the Department of Mechanical Engineering, University of Tokyo, in 1986, an associate professor, in 1989, and a professor in the Department of Mechano-Informatics, in 2000. His research interests include key technologies of robotic systems and software architectures to advance robotics research. His research projects have included hand-eye coordination in rope handling, visionbased robotic server systems, remote-brained robot approach, whole-body behaviors in humanoids, robot sensor suits with electrically conductive fabric, and developmental software systems for evolving a robot body with a spine. He has received several awards including outstanding Paper Awards in 1987, 1998, and 1999 from the Robotics Society of Japan, JIRA Awards in 1994, ROBOMECH Awards in 1994 and 1996 from the Division of Robotics and Mechatronics of the Japan Society of Mechanical Engineers, and Fellow of the Robotics Society of Japan in 2008.
Currently, he is a Chair of the Department of IRT Systems Study, Information and Robot Technology Research Initiative, University of Tokyo.
\end{IEEEbiography}

\end{document}

%% file: 01-intro.tex
\section{Introduction}
\IEEEPARstart{A}{erial} robots, especially multirotor unmanned aerial vehicles (UAVs), have been rapidly improved for various applications in the last few decades. Aerial manipulation is one of the most popular fields, and various types of aerial manipulators have been developed \cite{aerial_manipulator_general}.
In \cite{helicopter}, one of the earliest \revise{aircraft} is a helicopter with a heavy and high-performance industrial robotic arm. However, adopting this robot for daily manipulation tasks is difficult \revise{owing} to its complexity. Therefore, multirotors with \revise{various types of} arms have been \revise{proposed}. 
First, \revise{multirotors with a single-link arm were} developed (e.g., \cite{rigid-arm},~\cite{rigid-arm2},~\cite{rigid-arm3},~\cite{rigid-arm4},~\cite{rigid-arm5},~\cite{rigid-arm6}).
\revise{During a flight, these aerial robots equipped with rigid arms can achieve an arbitrary 6D pose at the end-effector when the dynamics of the robots are fully actuated \cite{rigid-arm-6d},~\cite{rigid-arm-6d-2},~\cite{rigid-arm-6d-3}}.
\revise{Additionally, the aerial manipulation tasks of under-actuated aerial robots have also increased as the number of joints or arms increased} (e.g., \cite{5dof-arm},~\cite{dual-arm2}). In \cite{tilt-3dof}, \cite{6dof-arm}, and \cite{7dof-arm}, \remove{the} multirotors with the three, six, and seven degrees of freedom (DoF) arm were developed, respectively. Furthermore, multirotors with two and three arms \revise{were} developed in \cite{dual-arm} and \cite{triple-arm}. \secondRevise{To achieve an arbitrary 6D pose at the end-effector using these under-actuated aerial robots during hovering, dynamics of manipulators are desired to be fully actuated. Therefore, the joint number of robots must be at least six \cite{actuation-definition}.}
Using these \revise{aerial} manipulators, basic manipulation tasks, such as point contact (PC) \cite{tilt-3dof}, pick \& place (P\&P) \cite{aerial_task:P&P}, pull/pushing (PP) \cite{ODAR}, sliding (S) \cite{aerial_task:Slide}, \revise{and} peg-in-hole (PH) \cite{aerial_task:PH} \revise{can be} conducted. In addition, more complex tasks, such as opening a lightweight valve \cite{dual-arm} and \revise{performing} corrosion repairs \cite{aerial_task:Repair} \revise{were} conducted.

\begin{figure}[t]
  \begin{center}
  \vspace{2mm}
    \includegraphics[width=1.0\columnwidth]{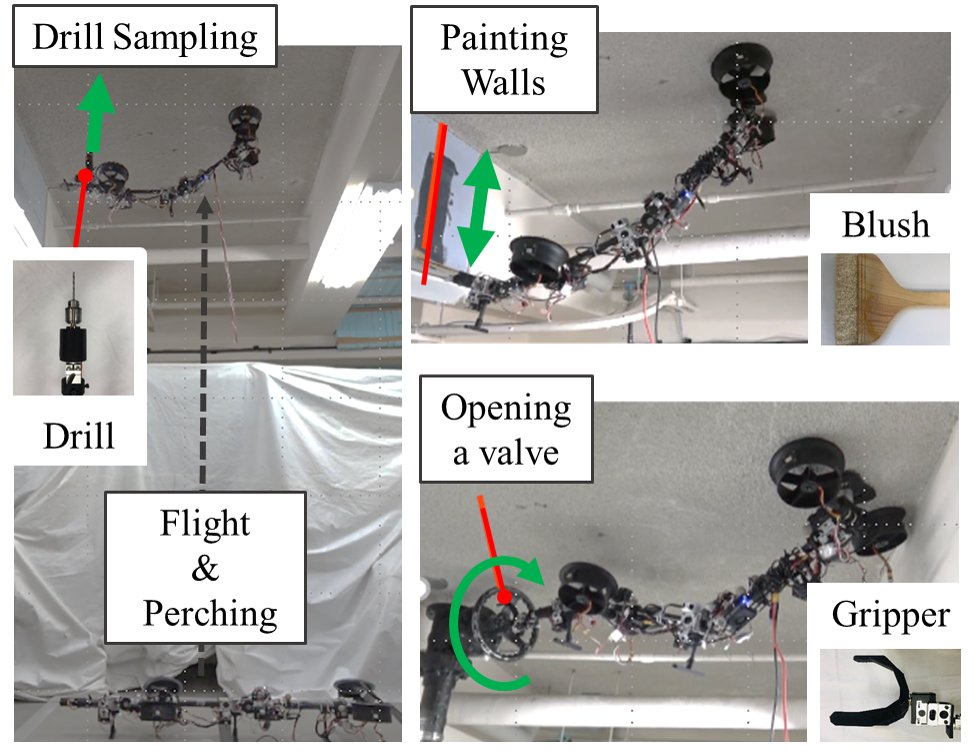}
  \end{center}
  \vspace{-3mm}
    \caption{\remove{The a}\add{A}erial manipulator can perch on horizontal surfaces using rotor thrust and perform manipulation tasks\add{,} such as drilling, painting, and opening \add{a} valve. \revise{(URL : https://www.youtube.com/watch?v=neKcLcekHMY)}}
  \label{figure:perching_arm_image}
\end{figure}

As mentioned above, the performance of the aerial manipulation has been significantly improved.
However, three critical problems need to be solved to improve their performance \revise{with regard to} more complex tasks:
(A) The end effector stability in the air is insufficient for more complex tasks. (B) \revise{Owing} to the payload capacity, the arm joint-torque limitations are very strict. Therefore, conventional aerial manipulators, i.e., multirotors with arms, can generate a comparatively small wrench at the end-effector.
(C) The end-effector motion is disturbed by the concentrated rotors. This causes \revise{a} reduction \revise{in the reachability of the end-effector}. 
To address these problems, we consider the minimal configuration of rotor-distributed manipulators that can perch on the environment. The perching ability significantly improves the end-effector stability. \revise{Thus,} the distributed rotors can relax the joint torque limitation and increase the reachability and feasible wrench at the end-effector. \revise{Consequently}, the proposed manipulator attains complex manipulations such as drilling, painting, and opening a valve, as shown in Fig.~\ref{figure:perching_arm_image}.

\subsection{Related Work}
\revise{Conventional aerial manipulators face three challenges}: (1) stability, (2) feasible wrench \revise{generation}, and (3) reachability at the end-effector. In the following sections, we introduce \revise{related work and our} approaches \revise{for} address these problems.
\subsubsection{Perching Approach to Increase the Stability}
A fixed-body approach, \revise{in which robots use an adhesive to perch on surface}, is one of the most useful approaches for aerial robots \revise{that can be used} to improve the precision at the end effector. 
Previous works have proposed various mechanisms to perch on \revise{various surfaces} (e.g.,~\cite{perching:cup},~\cite{perching:cup2},~\cite{perching:spine},~\cite{perching:spine2}).
A bio-inspired adhesive that is useful to perch on flat planes has been suggested (e.g.,~\cite{perching:adhesive},~\cite{perching:dry-adhesive}). This approach can significantly reduce their energy consumption \revise{as} the multirotor is not required to generate thrust during perching. However, this mechanism cannot exert a \revise{sufficient} force on rough planes. \revise{Another popular approach} is a gripper (e.g.,~\cite{perching:gripper}.~\cite{perching:gripper2},~\cite{perching:bird-like-gripper}). This approach is effective for perching on columnar objects such as pipes\revise{;} however, \revise{its mechanics make perching on planes difficult}. Magnetic perching is also a popular approach. \revise{It can exert comparatively large force only for magnetic metallic objects} (e.g.,~\cite{perching:magnet},~\cite{perching:magnet2})
\revise{Additionally, the} simplest perching approach for aerial robots \revise{is using} the rotor suction force. The benefit of this approach is that \revise{no additional modules are needed to increase the weight, such as magnets or grippers.} In \cite{perching:rotor-suction}, a quadrotor for bridge inspections was developed, and it can perch on ceilings using the rotor thrust. This \revise{study} also \revise{indicated} that the perching ability \revise{of robots} on ceilings is useful for aerial manipulation tasks, such as inspections and repairs. Here, the rotor suction force is varied near ceilings \revise{owing} to the airflow change, \revise{a phenomenon} called the ceiling effect. \revise{This effect} depends on the distance between the rotors and ceilings. Some studies have investigated \revise{this} ceiling effect (e.g.,~\cite{ceiling_effect_analysis},~\cite{ceiling_effect_analysis2},~\cite{ceiling_effect_analysis3},~\cite{ceiling_effect_analysis4}). Several models have been proposed based on the fluid dynamics (e.g.,~\cite{ceiling_effect_model},~\cite{ceiling_effect_model2}). 
In the conventional perching approaches using rotor thrust, the suction force is uncontrolled, and rotors generate excessive thrust. To address this problem, we propose a controller that calculates the minimum rotor thrust to perch. 

\subsubsection{Rotor-Distributed Manipulator to Increase the End-Effector Feasible Wrench}\label{sec1:subsec:RDM}
To address the small feasible wrench problem at the end effector, the rotor arrangement must be considered. By using distributed rotors on arm links, the rotor thrust can support the arm joint torque. \revise{Therefore, the} rotor-distributed manipulator (RDM) end-effector can generate a larger wrench than the conventional rotor-concentrated manipulators (RCM). Some aerial robots based on this rotor-distributed approach have been proposed already (e.g.,~\cite{flight-array},~\cite{flying-gripper},~\cite{hydrus},~\cite{jet-hr2}). 
In \cite{dragon}, a rotor-distributed aerial robot whose links are connected by two servo motors is developed. \revise{Each link} has a dual-rotor unit that generates equal thrust. These rotors' thrust relaxes \revise{the} joint torque limitations. Based on the quadrotor design, each rotor unit is placed at the center of a link. This robot can fly when the numbers of links and rotors are more than four and eight. By utilizing the rotor-distributed approach, the robot can achieve various postures in the air. In \cite{lasdra}, a rotor-distributed aerial robot with \revise{sixteen} rotors, whose links are connected by spherical passive joints and equipped with eight rotors, is proposed. \revise{The dynamics of each link} is fully actuated, and the rotor arrangement of each link is optimized considering the feasible control wrench as shown in \cite{ODAR}. Using the fix-body approach on the ground, this large-scale robot \revise{can complete} manipulation task, such as opening a valve. 
\revise{However}, the conventional rotor-distributed robots are \revise{large} and have several rotors, \revise{as} the \revise{end-effector} feasible wrench increases as the rotor number and diameter increase. However, it is difficult to use these \revise{large} robots instead of human arms in narrow spaces. 
Therefore, a human-scale arm robot that can enter narrow spaces \revise{and generate maximum possible wrenches} is required.
In this study, we propose a minimal configuration of \revise{a} human-arm-scale RDM\revise{, considering} the rotor number and arrangement based on flight stability and end-effector performance.

\subsubsection{Root-Perching to Increase Reachability}
Based on previous research, we consider the minimal configuration of \revise{a} human-arm scale RDM that can perch on ceilings using rotor suction force. To increase the reachability at the end-effector, the manipulator only uses a part of \revise{its} body for perching, such as \revise{the} root unit. Unlike the conventional RCM, the root unit of the RDM \revise{becomes} small, \revise{and its} perching stabilization becomes more difficult. Therefore, it is essential to consider the contact conditions, such as the static friction and zero moment point (ZMP) \cite{zmp}. Bipedal robots walk using \revise{the} rotor thrust considering the foot conditions \revise{to stabilize their walking} in \cite{jet-hr1} and \cite{leo}. In this \revise{study}, we propose a control method that minimizes the rotor thrust for perching, considering the foot conditions. Furthermore, we propose a motion planner and calculate the reachability of RDMs \revise{to evaluate their performance}. 

In this paper, we introduce the design, control, and motion planning methods for \revise{an} RDM that can perch on ceilings and evaluate the experimental performance.
The key contributions of this paper are as follows:

\begin{enumerate}
    \item We design \revise{a} minimal RDM with \revise{the root perching abilities} considering flight stability and end-effector performance. Additionally, we \revise{compare its} performance compared to the conventional RCM \revise{that is} composed of identical elements. 
    \item We \revise{propose} a controller to stabilize perching motions. The perching controller considers contact conditions, such as static friction and ZMP.
    \item Finally, we propose a motion planning method for the perching RDM considering some constraints such as perching force. This robot attains stable manipulation tasks on actual ceilings.
\end{enumerate}

\subsection{Organization}
In the following sections,
we introduce a human-arm-scale RDM design, control, and motion planning method. We discuss the minimal design of the RDM in Sec.~\ref{sec:design}. In Sec. \ref{sec:model}, the quasi-static model of the proposed RDM is derived. Based on the model, we propose a flight/perching control and motion planning method in Sec\revise{s}. \ref{sec:control} and \ref{sec:planning}. We introduce the system architecture in Sec.~\ref{sec:system} and evaluate the flight and perching motions in Sec. \ref{sec:experiment}, Finally, the results and future work are discussed in Sec.~\ref{sec:conclusion}. \revise{The main physical variables in this paper are defined in Table. \ref{table:variable-definition}.}

\begin{table}[t!]
 \caption{\revise{Definition of the main physical variables.}}
 \vspace{2mm}
 \label{table:variable-definition}
 \centering
  \begin{tabular}{cc}
   \hline
   \revise{Symbol} & \revise{Definition}\\
   \hline \hline
   \revise{$c$} & \revise{Ratio of the rotor thrust to its drag}\\
   \revise{$D_{rotor}$} & \revise{Rotor diameter}\\
   \revise{$\bm{F}$} & \revise{Force vector}\\
   \revise{$g$} & \revise{Gravity acceleration}\\
   \revise{$H$} & \revise{Foot size}\\
   \revise{$\bm{I}$} & \revise{Moment of inertia}\\
   \revise{$\bm{J}$} & \revise{Jacobian matrix}\\
   \revise{$m$} & \revise{Mass}\\
   \revise{$L$} & \revise{Length}\\
   \revise{$\bm{M}$} & \revise{Moment vector}\\
   \revise{$\bm{p}$} & \revise{Position vector}\\
   \revise{$\bm{q}$} & \revise{Joint vector}\\
   \revise{$\bm{Q}$} & \revise{Thrust allocation matrix}\\
   \revise{S} & \revise{Area}\\
   \revise{$\bm{V}$} & \revise{Volume}\\
   \revise{$\bm{W}$} & \revise{Wrench vector}\\
   \revise{$\alpha$} & \revise{Vectoring apparatus angle in the roll direction}\\
   \revise{$\beta$} & \revise{Vectoring apparatus angle in the pitch direction}\\
   \revise{$\eta$} & \revise{XYZ-Euler angle vector}\\
   \revise{$\theta$} & \revise{Roll angle}\\
   \revise{$\mu$} & \revise{Static friction coefficient}\\
   \revise{\bm{$\zeta$}} & \revise{Pose and joint angles vector}\\
   \revise{$\sigma$} & \revise{Rotational direction of a rotor}\\
   \revise{$\bm{\lambda}$} & \revise{Rotor thrust vector}\\
   \revise{$\bm{\xi}$} & \revise{Position and attitude vector}\\
   \revise{$\phi$} & \revise{Pitch angle}\\
   \revise{$\bm{\tau}$} & \revise{Joint torque vector }\\
   \revise{$\psi$} & \revise{Yaw angle}\\
   \revise{$\bm{\omega}$} & \revise{Angular velocity vector}\\
   \hline
 \end{tabular}
\end{table}

%% file: 02-design.tex
\section{Design}\label{sec:design}
This section introduces the minimal configuration of the human-arm scale RDM. The proposed manipulator is composed of an arm and foot unit to perch on ceilings. Herein, (A) some basic RDMs are introduced, and the number and arrangement of \revise{the} rotors are determined considering the flight and end-effector performance.
In addition, the proposed RDM is compared with the conventional RCM by focusing on the end-effector performance. (B) The detailed configuration of the proposed RDM is explained.

\subsection{Minimal Design of the Rotor-Distributed Manipulator}
To clarify the minimal configuration of the RDM, the number of joints $N_{joint}$, arm links $N_{link}$, and rotors $N_{rotor}$ need to be determined. The $N_{joint}$ is set to six. \revise{For aerial manipulators during flight, no arm joints are required to achieve an arbitrary 6D pose at the end-effector when the robot dynamics are fully actuated. However, for fixed-base RDMs, six joints are required to achieve arbitrary 6D pose at the end-effector. Here, the maximum DoF of a single joint unit is three such as the ball joint. Therefore, the minimum number of arm links $N_{link}$ for fixed-base RDMs is two such as LASDRA \cite{lasdra}. However,} the airflow caused by the end-effector rotor can \revise{disturb} the manipulation. \revise{Furthermore, the reachability space of these two-link RDM is not sufficient considering the rotor unit interference.}
Therefore, the \revise{link number for the proposed RDM} is set to three, and the end-effector link \revise{is not} equipped with any rotors. Furthermore, we choose the link length to attain the human-scale arm as shown in Sec.~\ref{subsec:platform_experiment}. Based on the above discussion, the number and arrangement of rotors is discussed considering the flight and end-effector performance. 

\subsubsection{Flight Stability by the Number of Rotors}
The flight performance of RDMs mainly depends on the number of rotors $N_{rotor}$. Several studies have determined the rotor numbers and arrangement for aerial robots with a rigid body (\cite{ODAR} and \cite{omni}). However, \revise{with these approaches, it is} difficult to evaluate the performance of the proposed rotor-distributed robots, \revise{as} the rotor arrangement varies \revise{owing} to the joint-angle change. Therefore, we introduce a simple approach to evaluate the flight stability of RDM, called "the aerial zero moment point (aerial ZMP)", which is analogous to the conventional ZMP method \cite{zmp}. The conventional ZMP approach is used to control the robots on the ground. \revise{Robot} stability is guaranteed when the ZMP is within the support polygon area determined by foot positions. In steady state, the ZMP corresponds to the projected point of the center of gravity (CoG) $\bm{p}_{CoG}$. 

This study evaluated the flight stability of RDM in a steady state. We assumed that the thrust units of RDMs are perpendicular to the ground, and each rotor generates enough thrust and drag moment around its rotational axes. Then, the altitude \revise{$z$} and yaw \revise{$\psi$} control stability is guaranteed. The position in the horizontal direction \revise{$x,~y$} is also stabilized using the vectoring apparatus angles or roll/pitch angles \revise{$\phi,~\theta$} like \revise{in} conventional multi-rotor robots (e.g., bi-rotor, tri-rotor, quad-rotor). However, it is challenging for \revise{the} RDM to stabilize the roll/pitch angles only using vectoring apparatus angles \revise{owing} to the comparatively large modeling errors caused by link-joint angles. In previous \revise{studies} \cite{lasdra},~\cite{hydrus},~and \cite{dragon}, all \revise{the} RDMs \revise{used} rotor thrust to stabilize the roll/pitch angles.
Therefore, it is effective to stabilize the roll/pitch control of RDMs using rotor thrust. Here, this stability is evaluated by the arrangement of rotors and the position of the CoG.

\begin{figure}[!t]
\centering
\includegraphics[width=3.0in]{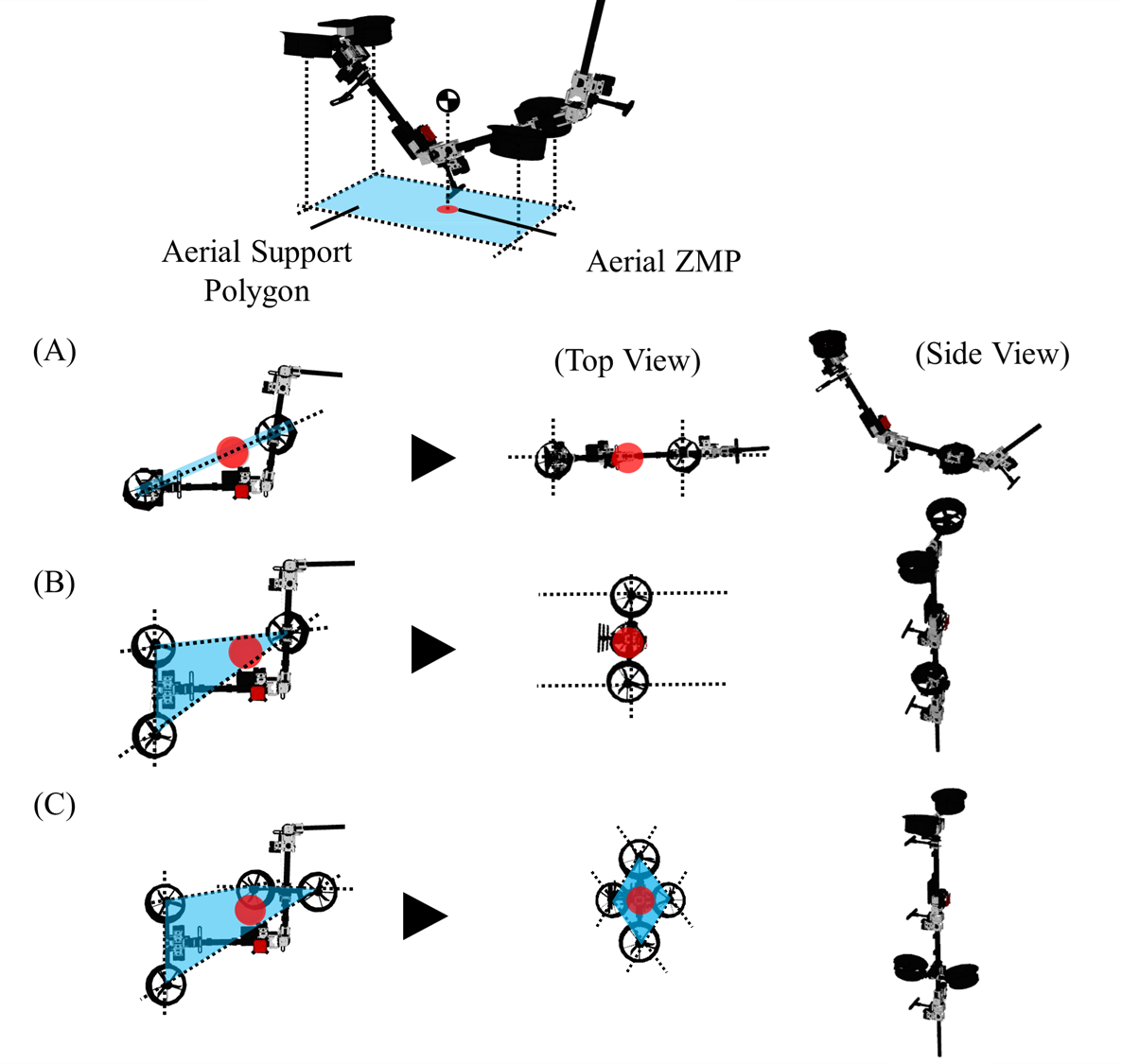}
\caption{Aerial zero moment point and support polygon area. (A) Bi-rotor flight stability. (B) Tri-rotor flight stability (C) Quad-rotor flight stability}
\label{fig:aerial_zmp}
\end{figure}

The aerial support polygon (ASP) area $S_{ASP}$ determined by the rotor positions is shown in Fig.~\ref{fig:aerial_zmp}~(A). This area is determined by

\begin{align}
\label{eq:aerial_support_polygon}
   S_{ASP}:=\{ 
\bm{p}_{ASP_{xy}}=\sum_{n=1}^{N_{rotor}}\gamma_i \bm{p}_{rotor,i_{xy}}| \gamma_i \geq 0, \sum_{n=1}^{N_{rotor}}\gamma_i=1\}
\end{align}

where the rotor position is $\bm{p}_{rotor,i_{xy}}\in \mathbb{R}^{2\times1}$.
Here, the flight stability is guaranteed when the aerial ZMP $\bm{p}_{AZMP_{xy}}\in \mathbb{R}^{2\times1}$, \revise{and} the projected point of the CoG, is within the ASP area in Eq.~\eqref{eq:aerial_support_polygon}. The detailed derivation is provided in Appendix \ref{appendix:aerial_stability}. 

\begin{align}
   \add{\bm{p}_{AZMP_{xy}} \in S_{ASP},}
   \label{eq:aerial_zmp}
\end{align}

In this study, we consider the basic design of RDMs in steady states, such as bi-rotor, tri-rotor, and quad-rotor \revise{RDMs}, as shown in Fig.~\ref{fig:aerial_zmp} (A)-(C). 
Each manipulator link is connected by joints. Each rotor unit can be turned upward by joints. 
We place one rotor unit on the tip of the root link for perching and another \revise{at} the center of the middle link. No rotors \revise{are placed} on the end link to \revise{prevent from} interfering with manipulations. 
In this study, the distance between rotors on the dual-rotor unit is as small as possible to reduce the joint load caused by the dual-rotor thrust difference, as shown in Fig\revise{s}.~\ref{fig:aerial_zmp}~(B) and (C).

To evaluate this RDMs stability, we focus on the configuration change. For the bi-rotor RDM, it is difficult to obtain a straight-line configuration during a flight \revise{as} the $S_{ASP}$ does not exist, as shown in Fig.~\ref{fig:aerial_zmp} (A).

Unlike conventional bi-rotors with \revise{rigid bodies}, actual RDMs \revise{cannot easily} fly using the change \revise{in the} rotor unit angles in the air \revise{owing} to modeling errors of \revise{the} link-joint angles. 
For the tri-rotor RDM, only the straight-line configuration flight that is perpendicular to the ground \revise{is} also difficult, \revise{as} the $S_{ASP}$ does not exist, as shown in Fig.~\ref{fig:aerial_zmp} (B).  
For the quad-rotor, a straight-line configuration flight that is perpendicular to the ground can be obtained since the aerial ZMP can be within the support polygon area with a margin, as shown in Fig.~\ref{fig:aerial_zmp} (C).
Therefore, we adopt the quad-rotor \revise{as} the minimal RDM configuration that can achieve any \revise{flight configuration}.

\begin{figure}[!b]
\centering
\includegraphics[width=3.0in]{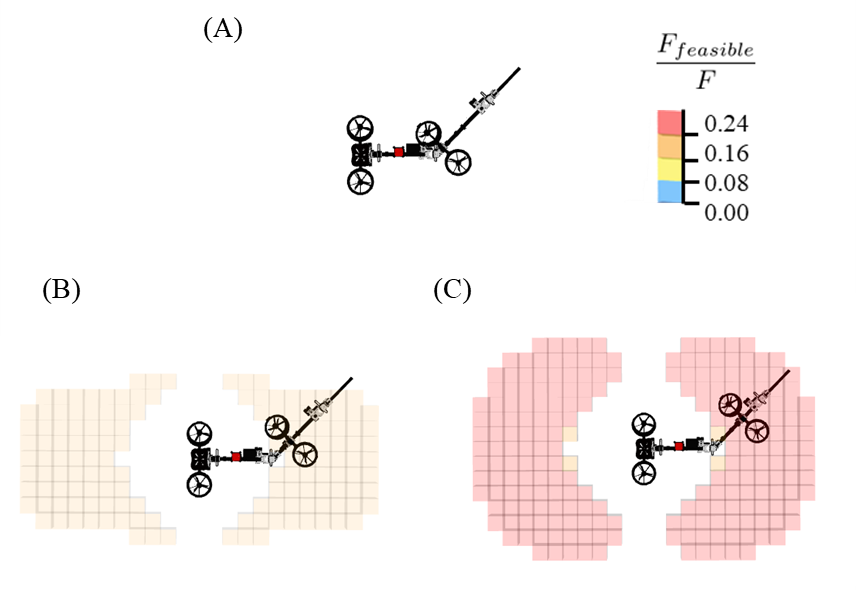}
\caption{Reachability due to the arm vectoring apparatus position: (A) root case; (B) middle case; and (C) end case.}
\label{fig:reachability}
\end{figure}

\subsubsection{End-Effector Performance by the Rotor Arrangement}
The arrangement of the rotor units is important for the reachability and feasible wrench at the end-effector. 
Therefore, a suitable value \revise{for evaluating} these performances \revise{needs} to be considered. Several approaches have been proposed to evaluate the manipulator performance. In \cite{manipulability}, the manipulability ellipsoid was proposed as an analytical evaluation value for manipulators.
This method can consider actuators' limitations such as joint torque using a weighting matrix \cite{dynamic_manipulability}.
In this study, we introduced a method \revise{for evaluating} the manipulator performance considering both feasible force and reachability at the end-effector. Here, the reachability space is calculated using inverse kinematics (IK) that considers constraints such as \revise{the} joint torque.
For RDMs, additional constraints such as rotor thrust and stability are considered. The aerial manipulator reachability (AMR) space $v_{AMR}$ and feasible force are calculated using the motion planning approach, as described in Sec.\ref{sec:planning}. Furthermore, the feasible force at the end effector $F_{ee, feasible}$ is considered. $F_{ee, feasible}$ is the maximum force that the end-effector is able to generate at a point in the reachability space.
 \begin{equation}
\add{F_{ee, feasible}=\max_{j}{F_{{ee, max}_j}}~(j=x,~y,~z)},
\label{eq:f_feasible}
\end{equation}
\add{where $F_{{ee, max}_j}=\max|F_{{ee}_j}|$.} 
Based on the above discussion, we defined an aerial manipulator feasibility (AMF) $v_{AMF}$ value that evaluates the reachability and feasible force at the end effector as follows:

 \begin{equation}
v_{AMF}=\dfrac{1}{L^3}\int_{\bm{V}_{AMR}}\dfrac{||F_{ee, feasible}||}{F}d\bm{V},
\label{eq:amf}
\end{equation}

where $F,~L$ are \revise{the} representative force and length of the aerial manipulator.
The $v_{AMF}$ is \revise{the} total value of the end-effector feasible force in the reachability space. It can evaluate both the end-effector reachability and feasible force.
Divided by these representative values $F,~L$, this non-dimensionalized AMF value can also be used to compare different lengths and weights \revise{manipulators' performance}.

\begin{table}[t!]
 \caption{AMF for varying arm vectoring-apparatus position.}
 \vspace{2mm}
 \label{table:manipulator_rotor_pos}
 \centering
  \begin{tabular}{cll}
   \hline
   Arm Vectoring Apparatus Position & AMF value \\
   \hline \hline
   Root & 0.0\\
   Middle & $1.5 \times 10^{-2}$\\
   End & $3.1 \times 10^{-2}$\\
   \hline
 \end{tabular}
\end{table}

A dual-rotor vectoring apparatus with two joints are placed at the tip of the root link for \revise{perching} on walls. For \revise{the} proposed quad-rotor, only the position of the arm vectoring apparatus on the middle link is determined, considering constraints such as rotor thrust limitations. The arm vectoring apparatus position is changed from the root to the end of the middle link to evaluate the end-effector performance as shown in Fig.~\ref{fig:reachability}. In the root case, $v_{AMF}=0.0$ \revise{as} the arm rotor thrust is insufficient to support the arm weight as shown in Fig.~\ref{fig:reachability} (A). Next, $v_{AMF}=1.5 \times 10^{-2}$ in the middle case and $v_{AMF}=3.1 \times 10^{-2}$ in the end case as shown in Table \ref{table:manipulator_rotor_pos}. These results indicate that the reachability and AMF value are \revise{maximums} in the end case. Therefore, we place the arm vectoring apparatus at the end of the middle link.

\subsubsection{Reachability Comparison}

\begin{table}[b!]
 \caption{Reachability and feasibility of\\ quadrotor-based RCM and RDM.}
 \vspace{2mm}
 \label{table:reachability_comparison}
 \centering
  \begin{tabular}{ccc}
   \hline
   Type & AMR [$m^2$] & AMF value \\
   \hline \hline
   RCM & 1.01 & $7.5 \times 10^{-2}$\\
   RDM & 2.86 & $3.2\times 10^{-1}$  \\
   \hline
   \multicolumn{2}{c}{Representative Froce} & 33.5 [N]\\
   \multicolumn{2}{c}{Representative Length} & 1.2 [m]\\
   \hline
  \end{tabular}
\end{table}

\begin{figure}[t!]
    \begin{center}
    \includegraphics[width=0.9\columnwidth]{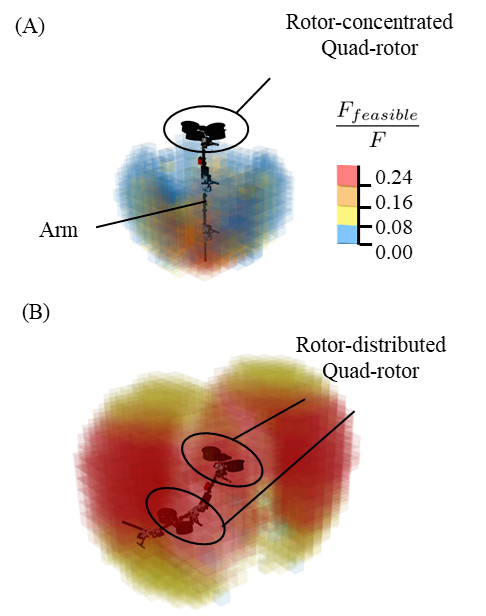}
    \caption{3-D reachability space: (A) RCM and (B) RDM}
    \label{fig:reachability_comparison}
    \end{center}
\end{figure}

This study aims to develop a minimal RDM whose performance exceeds conventional RCMs. 
Therefore, we compared a quadrotor-based RCM and RDM that are composed of identical elements \revise{as shown in Fig.~\ref{fig:reachability_comparison}}. As shown in Table \ref{table:reachability_comparison}, the AMR of RDM is 2.86 $\rm{m^2}$, which is approximately 2.8 times larger than \revise{that of the} RCM. Furthermore, the feasible end-effector force of RDM is also larger than \revise{that of} the RCM \revise{as} the distributed rotor thrust can relax the joint torque limitations. Therefore, the AMF value of RDM is $3.2\times 10^{-1}$ which is approximately forty times larger than \revise{that of the} RCM. 

\subsection{Platform}\label{subsec:platform}
\begin{figure}[t!]
    \begin{center}
    \includegraphics[width=0.9\columnwidth]{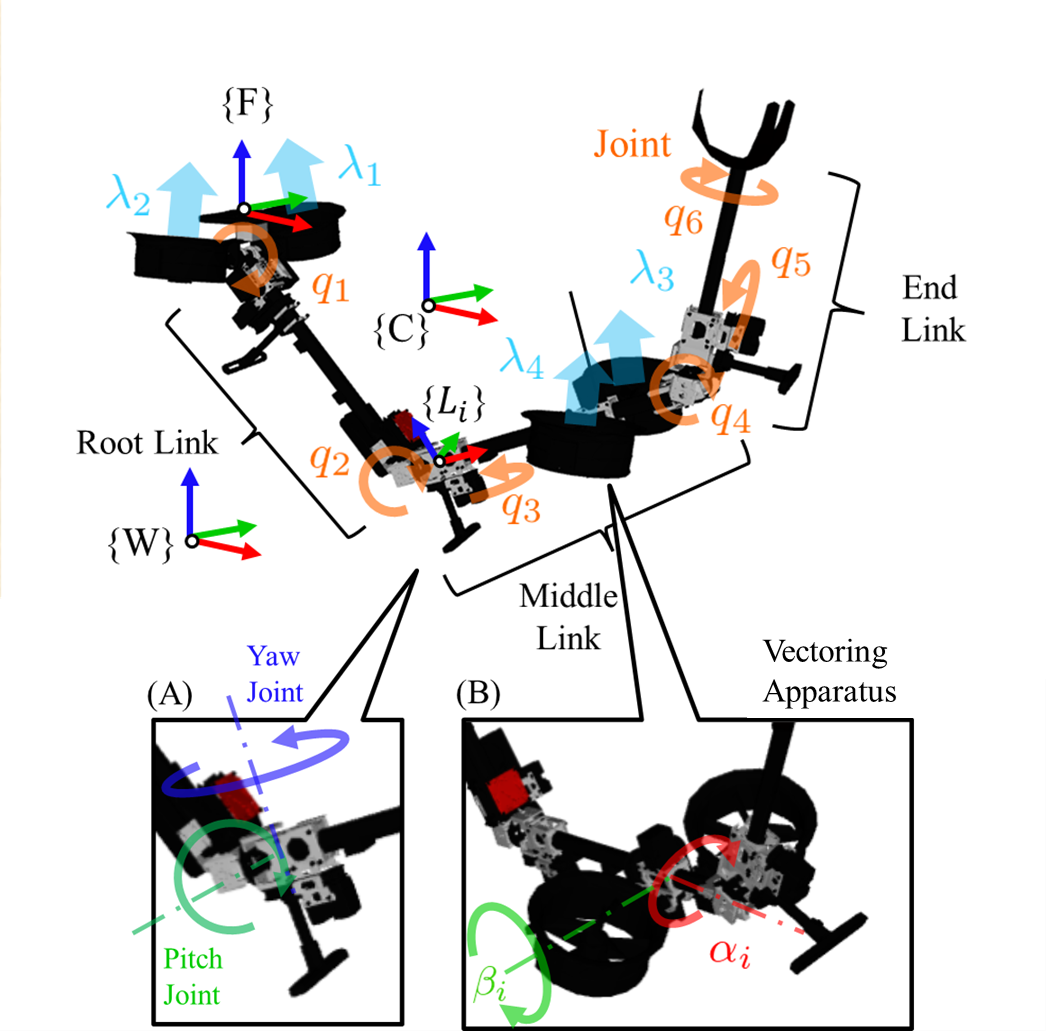}
    \caption{Robot model of the proposed rotor-distributed manipulator.}
    \label{fig:robot_model}
    \end{center}
\end{figure}

    The main frame link is cylindrical. Each link is connected by two joints $q_{2i},~q_{2i+1}(i=1,2)$ in the pitch and yaw directions as shown in Fig.~\ref{fig:platform} (A). These joints are changed from -90 to 90 deg. Here, the sum of the joint numbers is six, \revise{as} the root pitch $q_1$ and end-effector roll $q_6$ joints are equipped.
    The vectoring apparatus of the arm link has two joints in the roll and pitch directions $\alpha_i,~\beta_i$ to turn the thrust direction upward as shown in \revise{Fig.~\ref{fig:robot_model}} (B). These joints can be changed from -180 to 180 deg. 
    Here, the distance between the two rotors on a vectoring apparatus is as small as possible to reduce the load of the vectoring apparatus around the roll joint. We denote each rotor thrust $\lambda_i~(i=1,2,3,4)$.
The foot vectoring apparatus is placed at the tip of the root link to perch on ceilings. The footplate is approximated to the rectangle that has a wrench sensor to estimate the contact conditions. In addition, the end-effector unit can be replaced by a drill, brush, or gripper.
The processor and control board are placed at the tip of the root link to increase the end-effector feasible force. The software system is explained in Sec.\ref{subsec:platform_experiment}.

In this section, we introduced the quadrotor-based RDM with three links and six joints, considering the aerial ZMP for flight stability evaluation. 
To improve the manipulation performance, we determine the rotor arrangement considering the reachability and feasible force at the end-effector. Furthermore, we demonstrate that the proposed RDM performance significantly exceeds the conventional RCM composed of identical components. Finally, we explain the detailed configuration of the proposed manipulator. 

%% file: 03-model.tex
\section{Model}\label{sec:model}
In this section, we derive a model for flight and perching control of RDMs. 
The following coordinate systems
and notation conventions are used to describe the model. We
use four coordinate frames as shown in Fig.~\ref{fig:robot_model}: the world frame \revise{$\mathscr{W}$}, foot
frame \revise{$\mathscr{F}$}, CoG frame \revise{$\mathscr{C}$}, and i-th link frame \revise{$\mathscr{L}_{\mathscr{I}}$}. The world frame is
fixed on the ground with its z-axis pointing upwards. The foot frame is located on the footplate, its origin is at the
center of the footplate, and its y-axis corresponds to the rotational axis of the root joint. In addition, the CoG frame is located at the center of mass, and its x and y axes are aligned with those in the foot frame. Finally, each link frame origin is on the tip of the i-th main link. Their x-axes correspond to the respective axial direction of the i-th main frame. 
Furthermore, we use the following notation conventions: the
vector \revise{$^{}_{\mathscr{W}}\bm{r}$} represents the vector $\bm{r}$ in coordinate system \revise{$\mathscr{W}$}. In addition, the matrix \revise{$^{\mathscr{X}}_{\mathscr{Y}}R$} denotes the rotational matrix that rotates a vector in the coordinate system \revise{$\mathscr{X}$} to \revise{$\mathscr{Y}$}.

\subsection{Quasi-Steady Model for \revise{the} Rotor-Distributed Manipulator} For the modelling, we use the Newton-Euler formalism. The model of an articulated robot with distributed rotors is described as follows:

\begin{align}
   \begin{bmatrix}
   \bm{H}_{11} &\bm{H}_{12}\\
   \bm{H}_{21} & \bm{H}_{22}\\
   \end{bmatrix}
  \begin{bmatrix}
   \bm{\revise{\ddot{\xi}}} \\
   \bm{\ddot{q}} \\
   \end{bmatrix}
  +
   \begin{bmatrix}
   \bm{l} _{pose}\\
   \bm{l}_{joint} \\
   \end{bmatrix}
 =
    \begin{bmatrix}
   \bm{O_{6 \times 1}}\\
   \bm{\tau}_{joint} \\
   \end{bmatrix}
   +
   \bm{J}_{rotor}^\mathrm{T}
   \begin{bmatrix}
   \bm{F} _{rotor}\\
   \bm{M} _{rotor} \\
   \end{bmatrix}
   \nonumber\\ 
    \!\!\!\!\!\!\!\!\!+\!\!\!\!\!\!
  \sum^{N_{contact}}_{i=1}\!\!\!\bm{J}_{contact,i}^\mathrm{T}
   \begin{bmatrix}
   \bm{F} _{contact_i}\\
   \bm{M} _{contact_i} \\
   \end{bmatrix}
    +\!\!\!
  \sum^{N_{part}}_{i=1}
    \bm{J}^T_{part_i}(\bm{q})
    m_{part_i}\bm{g},
    \label{eq:general_model}
\end{align}
where $\bm{\revise{\ddot{\xi}}}=[\revise{\bm{\ddot{p}}^\mathrm{T}},~\bm{\dot{\omega}}\revise{^\mathrm{T}}]^\mathrm{T}$. Here, $\revise{\bm{p}}\in \mathbb{R}^{3 \times 1}$ and $\bm{\omega}\in \mathbb{R}^{3 \times 1}$ represent the position and angular velocity of the CoG, respectively. $\bm{q}\in \mathbb{R}^{N_{joints} \times 1}$ is the vector of \revise{the} joint and vectoring apparatus angles.
The second term on the left side of the equation represents the effects of \revise{the} Coriolis and centrifugal forces. 
On the right side, $\bm{\tau}_{joint}\in \mathbb{R}^{N_{joints} \times 1}$ is the joint torque. $\bm{F}_{rotor}\in \mathbb{R}^{3 \times 1}$ and $\bm{M}_{rotor}\in \mathbb{R}^{3 \times 1}$ in the second term represent the force and moment, respectively, caused by the rotors' thrust.
In addition, $\bm{F}_{contact_i}\in \mathbb{R}^{3 \times 1}$ and $\bm{M} _{contact_i}\in \mathbb{R}^{3 \times 1}$ are \revise{the} contact force and moment at the i-th contact part. The final term is the effect caused by gravity. Here, $m_{part,~i}$ is the mass of the i-th part, and $\bm{g}$ is the gravity acceleration \revise{vector}.
Furthermore, $N_{contact}$ and $N_{part}$ are the number of contact points and robot parts. $\bm{J}$ is the Jacobian matrix of each component. 
From Eq.~\eqref{eq:general_model}, we derive a robot model for the RDMs. Assuming that the maneuvering with the joint motion of the manipulator is in a quasi-steady state ($\ddot{\bm{q}}\approx\bm{0}$ and $\dot{\bm{q}}\approx\bm{0}$), the manipulator translational motion \revise{at} the $\mathscr{W}$ coordinate and rotational motion \revise{at} the $\mathscr{C}$ coordinate are given by  

\begin{align}
   \begin{bmatrix}
   m\bm{E} &\bm{O}^{3\times3}\\
   \bm{O}_{3\times3} & \bm{I}(\bm{q})\\
   \end{bmatrix}
  \begin{bmatrix}
   \revise{\bm{\ddot{p}}} \\
   \bm{\dot{\omega}} \\
   \end{bmatrix}
  \!\!+\!\!
   \begin{bmatrix}
   \bm{O}^{3\times1}\\
   \bm{\omega}\times\bm{I}(\bm{q})~\bm{\omega} \\
   \end{bmatrix}
   ~~~~~~~~~~~~~~~~~~~~~~~~~~~~~
   \nonumber\\  
 ~~
 =
   \begin{bmatrix}
   \bm{F} _{rotor}(\bm{\lambda}_{rotor},\bm{q})\\
   \bm{M} _{rotor}(\bm{\lambda}_{rotor},\bm{q})\\
   \end{bmatrix}
   \!\!+\!\!
   \begin{bmatrix}
   \bm{F} _{foot}\\
   \bm{M} _{foot} \\
   \end{bmatrix}
   \!\!+\!\!
   \begin{bmatrix}
   \bm{F}_{ee}\\
   \bm{M}_{ee} \\
   \end{bmatrix}
    \!\!+\!\!
   \begin{bmatrix}
   m\bm{\bm{g}}\\
   \bm{O}^{3\times1}
   \end{bmatrix},
   \label{eq:model_quasi_steady}
\end{align}
where $m$ and $\bm{I}(\bm{q})$ are the robot mass and moment of inertia, respectively. $\bm{F}_{rotor}$ and $\bm{M}_{rotor}$ are the force and moment caused by rotor thrust $\bm{\lambda}_{rotor}\in\mathbb{R}^{N_{rotor} \times 1}$ and joint angles, respectively.
$\bm{F}_{foot}$,~$\bm{M}_{foot}$ and $\bm{F}_{ee}$,~$\bm{M}_{ee}$ are the force and moment on the footplate and at the end-effector, respectively, and $\bm{g}=[0,~0,~-g]^\mathrm{T}$ \revise{using gravity acceleration $g$}. In Eq.~\eqref{eq:model_quasi_steady}, the relationship between the wrench caused by rotor \revise{$\bm{W}_{rotor}=[\bm{F} _{rotor}^{\mathrm{T}},~\bm{M} _{rotor}^{\mathrm{T}}]^{\mathrm{T}}$} and $\bm{\lambda}_{rotor}$ is nonlinear. Therefore, the cost \revise{incurred} to calculate the desired rotor thrust is high. To reduce this cost, we assume that the vectoring apparatus is horizontal in the world coordinate. Consequently, $\bm{W}_{rotor}(\bm{q})$ in the $\mathscr{C}$ coordinate is expressed by

\begin{eqnarray}
   \bm{W}_{rotor}(\bm{q})=\bm{Q}(\bm{q})\bm{\lambda}_{rotor},
\label{eq:rotor_moment}
\end{eqnarray}
where the rotor allocation matrix $\bm{Q}(\bm{q})$ is given by:

\begin{eqnarray}
   \!\bm{Q}(\bm{q})&=&
  \begin{bmatrix}
  \bm{u}_1(\bm{q})&\cdots&\bm{u}_i(\bm{q})&\cdots&\bm{u}_{N_{rotor}}(\bm{q})\\
  \bm{v}_1(\bm{q})&\cdots&\bm{v}_i(\bm{q})&\cdots&\bm{v}_{N_{rotor}}(\bm{q})
  \end{bmatrix},\label{eq:rotor_allocation}\\
  \bm{v}_i(\bm{q})&=& ([\bm{p}_{rotor_i}(\bm{q}) \times] -c \sigma_i )\bm{u}_i(\bm{q}),
\end{eqnarray}
where $\bm{u}_i(\bm{q})$ is the unit normal on the rotational plane\revise{,} and $\sigma_i(=(-1)^{i+1})$ is the rotational direction of the i-th rotor. Thus, $c$ is the ratio of the rotor thrust to its drag \revise{around} the rotational axis. Based on the above model, we calculate the desired rotor thrust during a flight as described in Sec.~\ref{subsec:flight-control}.

\subsection{Contact Stability Constraints}
For stable perching, the additional constraints at each contact point should be considered. To prevent slipping on walls, we consider the Coulomb friction constraints. Then, a contact force $\bm{F} _{contact_i}$ satisfies the following relationships:

\begin{align}
   \bm{F}_{contact_i} \cdot \bm{n}_{contact_i} > 0, \\
   ||\bm{F}_{contact_i} - (\bm{F}_{contact_i} \cdot \bm{n}_{contact_i} ) ~\bm{n}_{contact_i}||\nonumber\\
   \leq \mu_j (\bm{F}_{contact_i} \cdot \bm{n}_{contact_i}),\label{eq:static_friction}
\end{align}
where $\bm{n}_{contact_i}$ is a unit normal to the contact surface, and $\mu_j~(j=x,y)$ is the static friction coefficient. Approximating the contact wrench cone (CWC) in Eq.~\eqref{eq:static_friction} to a pyramid, the foot force $\bm{F}_{foot}$ satisfies the following conditions:

\begin{eqnarray}
   F_{foot_z}&>&0,
   \label{eq:sec4:floor_reaction}\\
   |F_{foot_j}|&\leq& \mu_j~F_{foot_z}~(j=x,~y).
   \label{eq:static_friction_approx}
\end{eqnarray}

In addition, we should prevent rotation during perching. Considering the rotational static friction, the constraint is given by 

\begin{align}
    ||M_{contact_i,z}|| ~~~~~~~~~~~~~~~~~~~~~~~~~~~~~~~~~~~~~~~~~~\nonumber\\ ~~~~<\!M_{contact_i, z}(\bm{F}_{contact_i},\bm{M}_{contact_i}),
    \label{eq:static_rotation_friction}
\end{align}
where $M_{contact_i, z}(\bm{F}_{contact_i},~\bm{M}_{contact_i})$ is the maximum rotational static friction moment. Assuming $M_{contact_i, z}(\bm{F}_{contact_i},~\bm{M}_{contact_i})=\mu_z F_{contact_i,z}$ in Eq.~\eqref{eq:static_rotation_friction}, the condition is written as follows:  

\begin{eqnarray}
   |M_{foot_z}| \leq \mu_z ~F_{foot_z},
\label{eq:rotating_prevention_approx}
\end{eqnarray}
where $\bm{M}_{foot}$ is the foot moment, and $\mu_z$ is the rotational static friction coefficient.
For the surface contact, the ZMP $\bm{p}_{ZMP_{i_j}}~(j=x,y)$ at the i-th contact surface should be within the support polygon area $S_i$ to prevent the i-th foot plate from peeling from the surface. Then, the constraint is given by 

\begin{align}
    p_{ZMP_{i_x}},~p_{ZMP_{i_y}}  \in S_i.
    \label{eq:zmp_constraints}
\end{align}

Here, the relationship between the ZMP and the contact wrench is given by

\begin{align}
    \bm{p}_{ZMP_i} \times \bm{F}_{contact_i} + \bm{M}_{contact_i} = \bm{0}, 
    \label{eq:zmp}
\end{align}
where $\bm{p}_{ZMP_i}=[p_{ZMP_i,~x},~p_{ZMP_i,~y},~p_{ZMP_i,~z}]$. Using the boundary points $\bm{p}_{bound_j}(j=1,\cdots,N_{bound})$ in the foot plate area $S_{foot}$, the ZMP constraints in Eq.~\eqref{eq:zmp_constraints} are expressed by

\begin{eqnarray}
   p_{ZMP_i,~x},~p_{ZMP_i,~y} ~~~~~~~~~~~~~~~~~~~~~~~~~~\nonumber\\~~~~~~~~~~~\in \{\!\!\sum^{N_{bound}}_{j=1} \!\!\! \gamma_j \bm{p}_{bound_j} |  \bm{p}_{bound_j} \in S_{foot} \},
   \label{eq:zmp_constraint_approx}
\end{eqnarray}
where $N_{bound}$ is the number of boundary points. $\gamma_i$ satisfies the following constraints:

\begin{eqnarray}
    \begin{cases}
        {~~\gamma_j \geq 0}\\
        {~~\sum^{N_{bound}}_{j=1} \gamma_j = 1}.
    \end{cases}
   \label{eq:sec4:zmp_constraint}
\end{eqnarray}

Considering these constraints for calculating the rotor thrust, we achieve stable perching control as described in Sec.~\ref{subsec:perching-control}.

In this section, we introduce \revise{a} model of rotor-distributed robots in \revise{a} quasi-steady state and \revise{under} contact stability constraints, such as static friction and ZMP. Using these models, we propose a control method of the proposed minimal RDM for flight and perching.

%% file: 04-control.tex
\section{Control}\label{sec:control}
Our goal is to determine the rotor thrust and vectoring apparatus angles. Here, the desired rotor thrust is expressed as follows:

\begin{eqnarray}
\bm{\lambda}_{rotor}=\bm{\lambda}_{flight}+\bm{\lambda}_{perch},
\end{eqnarray}
where $\bm{\lambda}_{flight}$ is the thrust for flight, and $\bm{\lambda}_{perch}$ is the additional thrust for perching.
To decrease the computational cost, we use a linearized model and propose \revise{a controller for under-actuated robots} based on the back-stepping control framework. This method for minimal RDMs would be adopted to the RDMs with an increased number of links and rotors. The variable with the notation tilde represents the estimated value.
\begin{figure*}[t!]
    \begin{center}
    \includegraphics[width=1.95\columnwidth]{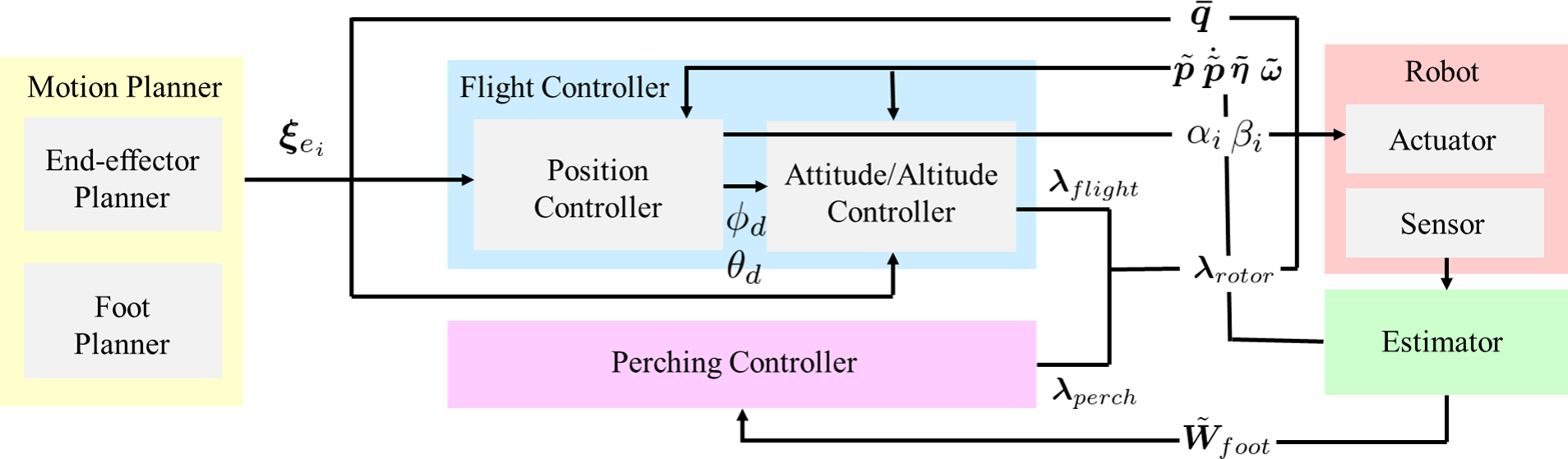}
    \caption{\revise{Control} flow for \revise{the} proposed perching manipulator}
    \label{fig:control_architecture}
    \end{center}
\end{figure*}

\subsection{Flight Control}\label{subsec:flight-control}
To achieve stable maneuvering with joint motion during flight, we propose a new control method that determines rotor thrust, attitude, and vectoring apparatus angles based on the conventional under-actuated control method. In the following sections, we introduce (1) attitude and altitude control and (2) position control in a horizontal plane.
\subsubsection{Attitude and Altitude Control}
To stabilize the attitude and altitude, the rotor thrust $\bm{\lambda}_{flight}$ is calculated using the linear quadratic integral (LQI) control method \cite{hydrus}. From Eq.\eqref{eq:model_quasi_steady}, the linearized equation for altitude and attitude is given by

\begin{eqnarray}
\label{eq:control_model}
\bm{\dot{x}}&=&\bm{Ax}+\bm{B}(\bm{q})\bm{\lambda}\add{_{flight}}+\bm{D},\nonumber\\
\bm{y} &=& \bm{Cx}.
\end{eqnarray}

Here, each matrix in Eq.~\eqref{eq:control_model} is expressed by

\begin{equation}
  \bm{x}=
    \begin{bmatrix}
      \bar{\bm{\xi}}\\
      \bar{\dot{\bm{\xi}}}
    \end{bmatrix},~
\bm{A}=
    \begin{bmatrix}
      \bm{O}^{4\times4}&\bm{E}^{4\times4}\\
      \bm{O}^{4\times4}&\bm{O}^{4\times4}\\
    \end{bmatrix},\nonumber
     \bm{B}=
    \begin{bmatrix}
      \bm{O}^{4\times N_{rotor}}\\
      \bar{\bm{M}}^{-1} \bar{\bm{Q}}(\bm{q})
    \end{bmatrix},\nonumber
\end{equation}

\begin{equation}
\bm{C}=
    \begin{bmatrix}
    \bm{E}^{4\times4}\\
    \bm{O}^{4\times4}\\
    \end{bmatrix},
\bm{D}=
    \begin{bmatrix}
   \bm{O}^{4\times 1} \\
   -g\\
   \remove{\bm{O}^{3\times 1}}\add{-\bm{I}^{-1}(\bm{q})(\bm{\omega}\times\bm{I}(\bm{q})~\bm{\omega})}
   \end{bmatrix},
\end{equation}

\begin{equation}
\bar{\bm{M}}=
    \begin{bmatrix}
   m & \bm{O}^{1\times3} \\
   \bm{O}^{3\times1} & \bm{I}(\bm{q})
   \end{bmatrix},\nonumber
   \label{eq:control_model_components}
\end{equation}
where $\bm{\bar{\xi}}$ and $\bm{\bar{Q}}(\bm{q})$ are the third to sixth rows of $\bm{\xi}$ and $\bm{Q}(\bm{q})$.

\remove{
\begin{equation}
  \bm{\lambda}_{comp}=\bar{\bm{Q}}(\bm{q})^{-1}
  \begin{bmatrix}
      0\\
      \bm{\omega}\times\bm{I}(\bm{q})~\bm{\omega}
    \end{bmatrix},
\end{equation}}

Note that $\bm{W}_{foot}=\bm{O^{6 \times 1}}$ in Eq.\eqref{eq:control_model} during a flight.
For the LQI control, we define the following values.

\begin{eqnarray}
    \label{eq:lqi_state_input}
    \overline{\bm{\xi}}&=&\bm{\xi}-\bm{\xi}_s,\\
    \overline{\bm{\lambda}}\add{_{flight}}&=&\bm{\lambda}\add{_{flight}}-\bm{\lambda}_{\add{{flight}}, s},
\end{eqnarray}
where $\bm{\xi}_s$ and $\bm{\lambda}_{\add{{flight}}, s}$ are steady values. 
When we define the tracking error $\bm{e}=\bm{x}-\bm{x}_s$, the integral value $\bm{v}$ is given by

\begin{eqnarray}
 \dot{\bm{v}} = \bm{y}_s-\bm{y} = -\bm{Ce}.
 \label{eq:integral_term}
\end{eqnarray}

From Eqs.\eqref{eq:control_model},~\eqref{eq:lqi_state_input},~and \eqref{eq:integral_term}, the modified equation is expressed as follows:

\begin{equation}
\dot{\bar{\bm{x}}}=\bm{\overline{A}~\overline{x}}+\bm{\overline{B}(\bm{q})~\overline{\lambda}}\add{_{flight}},
\end{equation}

\begin{equation}
  \overline{\bm{x}}=
    \begin{bmatrix}
    \bm{e}\\
    \bm{v}\\
    \end{bmatrix} ,~ 
    \overline{\bm{A}}=
    \begin{bmatrix}
   \bm{A} & \bm{O}_{8\times4}\\
   -\bm{C} & \bm{O}_{4\times4}\\
   \end{bmatrix},
  \overline{\bm{B}}(\bm{q})=
    \begin{bmatrix}
   \bm{B}(\bm{q}) \\
   \bm{O}_{4\times N_{rotor}}\\
   \end{bmatrix}.\nonumber
\end{equation}


The cost function is defined as follows:

\begin{gather}
    \bm{J} = \int^{\infty}_{0}(\overline{\bm{x}}^\mathsf{T}\bm{W}_1~\overline{\bm{x}}+\bm{\overline{\lambda}}^\mathsf{T}\add{_{flight}}\bm{W}_2~\overline{\lambda}\add{_{flight}})~dt,
    \label{eq:control_cost_func}
\end{gather}
where $\bm{W}_1$ and $\bm{W}_2$ are weight matrices.
By solving the algebraic Riccati equation (ARE), the optimal control input is obtained by

\begin{eqnarray}
    \bm{\lambda}\add{_{flight}}=\bm{K}_x \overline{\bm{x}} + \bm{\lambda}_{\add{flight}, s},
\end{eqnarray}
where $\bm{\lambda}_{\add{flight,}s}$ is used to support the gravity \add{and gyro moment} effect.
However, we can ignore the effect of gravity, \revise{as} the integral feedback term guarantees this effect. Therefore, the thrust for the flight is given by

\begin{eqnarray}
    \bm{\lambda}\add{_{flight}}=\bm{K}_x \overline{\bm{x}} + \remove{\bm{\lambda}_{comp}}\add{\bar{\bm{Q}}(\bm{q})^{-1}
  \begin{bmatrix}
      0\\
      \bm{\omega}\times\bm{I}(\bm{q})~\bm{\omega}
    \end{bmatrix}}.
    \label{eq:rotor_thrust_flight}
\end{eqnarray}

Using the thrust in Eq.~\eqref{eq:rotor_thrust_flight}, we control the attitude and altitude. Next, the position control method is introduced in a horizontal plane.

\subsubsection{Position Control in the Horizontal Plane}
In \revise{previous studies}, only the attitude in the roll and pitch directions were used for \revise{position control of under-actuated robots} in the horizontal plane \cite{hydrus}. However, this method is difficult to adopt for three-dimensional maneuvering with the joint motion of RDMs owing to the poor responsiveness caused by the significant moment of inertia. Therefore, both attitude and vectoring apparatus angles are used to control the $x$- and $y$- directions. Here, we assume that these angles are sufficiently small and do not affect the attitude and altitude control. 
Using the PID control, the desired acceleration $\ddot{\revise{p}}_{j,d}~~(j=x,~y)$ is given by

\begin{align}
  \remove{\ddot{\bm{\revise{p}}}_{d}}\add{\ddot{\revise{p}}_{j,d}}= k_{P_j} e_j\!+\!k_{I_j}\int \!e_j dt +k_{D_j} \dfrac{d}{dt}\dot{e_j} ~~(j=x,y),
  \label{eq:PID}
\end{align}
where $e_j$ is the state error in the world coordinate. Further, $k_{P_j}$, $k_{I_j}$,~and $k_{D_j}$ are the PID gains. Using the conventional approach, the aerial robot can be stabilized using the attitude in the roll and pitch angles as follows:

\begin{eqnarray}
    \phi_d &=& \add{k_{\phi_d}}\dfrac{\ddot{\revise{p}}_{x_d} \sin{\psi_d} - \ddot{\revise{p}}_{y_d} \cos{\psi_d} }{g},\\
    \theta_d &=& \add{k_{\theta_d}}\dfrac{\ddot{\revise{p}}_{x_d} \cos{\psi_d} + \ddot{\revise{p}}_{y_d} \sin{\psi_d}}{g}.
    \label{eq:attitude_angle}
\end{eqnarray}
During the flight of the proposed manipulator, \revise{it is desirable for the posture to approximate a straight, as} the stability and thrust margin become significant considering the aerial ZMP. In the above situations, it is challenging to stabilize the flight in the $x$ direction \revise{owing} to the response latency of the pitch angle, \revise{as} the moment of inertia in the pitch direction increases significantly. To address this problem, we set $\phi_d=0$ and use the i-th vectoring apparatus angles $\alpha_i$ and $\beta_i$ as follows:
        
\begin{eqnarray}
 \alpha_{i} &=& \tan^{-1}\left(\dfrac{-n_{\alpha_{i_y}}}{n_{\alpha_{i_z}}}\right),\\
 \beta_{i} &=& \tan^{-1}\left(\dfrac{n_{\beta_{i_x}}}{-n_{\beta_{i_y}}\sin\alpha_i+n_{\beta_{i_z}}\cos\alpha_i}\right),
\label{eq:gimbal_angle}
\end{eqnarray}
where the $n_{\alpha_{i}}$ and $n_{\beta_{i}}$ are expressed by

\begin{eqnarray}
    \bm{n}_{j_{i}}\!\!&=&\!\!\dfrac{\bm{n}_{va,j_i}}{||\bm{n}_{va,j_i}||}~~(j=\alpha,~\beta),\\
    \label{eq:gimbal_unit_normal}
    \bm{n}_{va,j_i}\!\!&=&\!\! ^{C}_{L_i}\!\bm{R}
    \begin{bmatrix}\!
    \remove{u_{mask_{j_i}}}\add{k_{\theta_i}}(\ddot{\revise{p}}_{x_d} \cos{\psi_d} + \ddot{\revise{p}}_{y_d} \sin{\psi_d})\\
    \remove{u_{mask_{j_i}}}\add{k_{\phi_i}}(\ddot{\revise{p}}_{x_d} \sin{\psi_d} - \ddot{\revise{p}}_{y_d} \cos{\psi_d}) \\
    g
    \end{bmatrix}\!,
\end{eqnarray}
where $\alpha_{i}$ cannot be used for position control \revise{owing} to the joint torque limitation \revise{as} the rapid \revise{changes} of the vectoring apparatus also generates the large torque on the roll joint in addition to the dual rotor thrust difference. Based on the above discussion\revise{, we} determine the above gains $k$ in experiments as shown in Sec.~\ref{sec:experiment}. Using the controller mentioned above, we stabilize the proposed RDM with a minimal configuration.

\subsection{Perching}\label{subsec:perching-control}
A method \revise{for perching} on surfaces is proposed to calculate the additional thrust for perching $\bm{\lambda}_{perch,t}$ at the time $t$. To minimize the thrust, we use the following framework.

\begin{align}
\label{eq:qp_opt_perch}
  \min_{\bm{\lambda}_{perch,t}} \bm{\lambda}_{perch,t}^\mathsf{T}\bm{N}\bm{\lambda}_{perch,t},
\end{align}

where $\bm{\lambda}_{perch,t}$ should be determined considering the foot contact conditions. To achieve stable perching, we define the perching thrust as follows:

\begin{eqnarray}
\label{eq:perching_thrust}
  \bm{\lambda}_{perch,t} = \bm{\lambda}_{perch,t}'+\bm{\lambda}_{perch,t}'',\nonumber~~~~~~~~~~~~~\\
  \bm{U}_{mask}(\bm{\bar{Q}}_{perch}(\bm{q})\bm{\lambda}_{perch,t}'-\bm{W}_{foot,t})=\bm{O},~~~~~~\\ \bm{U}_{mask}(\bm{\bar{Q}}_{perch}(\bm{q})\bm{\lambda}_{perch,t}''-\bm{W}_{foot,t-1}-\tilde{\bm{W}}_{foot})=\bm{O},\nonumber
\end{eqnarray}
where $\bm{\bar{Q}}_{perch}$ is the thrust allocation matrix for perching, and $\bm{U}_{mask}$ is the matrix \revise{used} to select the control wrench. Furthermore, $\bm{\lambda}_{perch,t}'$ is determined by considering the constraints of the foot conditions. $\bm{\lambda}_{perch,t}''$ is the feedback term that compensates the foot wrench errors.  
To increase end-effector stability, we use the force perpendicular to the contact plane that is generated by the foot rotors. 
Based on this strategy, we determine $\bm{W}_{foot,t}$ considering constraints such as static friction, moment, and ZMP.
From Eqs.~\eqref{eq:static_friction_approx} and \eqref{eq:rotating_prevention_approx}, the slipping prevention constraints in the foot coordinate are expressed by

\begin{eqnarray}
   F_{foot_z,t}&>&\epsilon_z,
   \label{eq:floor_reaction}\\
  |\tilde{F}_{foot_j}|&<&\mu_x~F_{foot_z,t}~(j=x,~y),\label{eq:sliding_prevention}\\
  |\tilde{M}_{foot_z}|&<&\mu_z~F_{foot_z,t},
   \label{eq:rotating_prevention}
\end{eqnarray}
where $\epsilon_z\add{(>0)}$ is \revise{a} sufficiently small offset value. To prevent peeling from surfaces, the ZMP should be within the footplate area. Assuming the support polygon area is rectangular, the ZMP constraints in Eq.~\eqref{eq:zmp_constraint_approx} can be represented by 
\begin{eqnarray}
   -H_j \leq p_{ZMP_j}\leq H_j~~(j=x,y),
   \label{eq:zmp_constraint_reqtangular}
\end{eqnarray}
where $H_j$ is the half-length of each rectangle edge. The ZMP can be expressed by

\begin{align}
     p_{ZMP_x} &=& \dfrac{\tilde{F}_{foot_x}\times p_{ZMP_z}-\tilde{M}_{foot_y}}{F_{foot_z,t}}\label{eq:zmp_x}\\
     p_{ZMP_y} &=& \dfrac{\tilde{F}_{foot_y}\times p_{ZMP_z}+\tilde{M}_{foot_x}}{F_{foot_z,t}}.\label{eq:zmp_y}
\end{align}

Additionally, the rotor thrust limitations must be considered and are as follows:

\begin{eqnarray}
    \bm{\lambda}_{min} \leq \bm{\lambda}_{perch, t}\leq \bm{\lambda}_{max},    \label{eq:rotor_thrust_constraint}
\end{eqnarray}
where $\bm{\lambda}_{min}$ and $\bm{\lambda}_{max}$ are the minimum and maximum limitations of \revise{the} rotor thrust\revise{, respectively}. These values are given by

\begin{eqnarray}
    \bm{\lambda}_{min}&=&\bm{O}^{N_{rotor}\times1},\\
    \bm{\lambda}_{max}&=&\bm{\lambda}_{rotor_{max}}+\bm{\bar{Q}}^{-1}(\bm{q})m\revise{\tilde{\bm{g}}},
    \label{eq:min_max_thrsut}
\end{eqnarray}
where $\bm{\lambda}_{rotor_{max}}$ is the maximum thrust for each rotor, and \revise{$\tilde{\bm{g}}=[-g, 0, 0, 0]^{\mathsf{T}}$}. 
Furthermore, the thrust must be changed continuously. Therefore, the following constraint is defined.
\begin{eqnarray}
    |\bm{\lambda}_{perch, t} - \bm{\lambda}_{perch, t-1}| \leq \delta\bm{\lambda}_{perch},
    \label{eq:continuous_thrust}
\end{eqnarray}
where $\delta\bm{\lambda}_{perch}$ is the limitation of the rotor thrust change at each step. From the above discussion, the perching thrust is calculated by the following quadratic programming (QP) problem.

\begin{eqnarray}
\label{eq:qp}
  \min_{\bm{\lambda}_{perch,t}} \bm{\lambda}_{perch,t}^\mathsf{T}\bm{N}\bm{\lambda}_{perch,t},~~~~~~~~~~~~~~~~~~ \\
  \rm{s.t.}~~~~~~~~~~~~~~~~~~~~~~~~~~~~~~~~~~~~~~~~~~~~\nonumber\\
  \rm{Eqs.}~\eqref{eq:perching_thrust},~\eqref{eq:floor_reaction},~\eqref{eq:sliding_prevention},~\eqref{eq:rotating_prevention},~\eqref{eq:zmp_constraint_reqtangular},~\eqref{eq:rotor_thrust_constraint},~\rm{and}~\eqref{eq:continuous_thrust}.\nonumber
\end{eqnarray}

Using the proposed perching control, we stabilize \revise{the motion of} the RDM motion on ceilings.

In this section, we propose a flight and perching controller to stabilize the RDMs. The flight position controller uses both attitude and vectoring apparatus angles. The QP-based flight controller considers foot conditions such as static friction and ZMP.

%% file: 05-planning.tex
\section{Planning}\label{sec:planning}
This section proposes a motion generation method for perching RDMs in the steady state. For applications, it is critical for the proposed RDM to achieve the desired end-effector pose. To generate the desired end-effector trajectories, the following framework is used based on differential kinematics:

\begin{gather}
\min_{\delta\bm{\zeta}_k}{G(\delta\bm{\zeta}_k)}, 
\label{eq:opt_plan}\\
   \bm{\zeta}_{k+1} = \bm{\zeta}_k + \delta \bm{\zeta}_k.
  \label{eq:update_state}
\end{gather}
First, $\delta\bm{\zeta}_k$ is determined to minimize the cost function $G(.)$ as shown in Eq.\eqref{eq:opt_plan}\add{.}\remove{, and update} $\bm{\zeta}_k$ is updated using $\delta\bm{\zeta}_k$ as shown in Eq.~\eqref{eq:update_state}. Here, the state variables are $\bm{\zeta}_k=[\bm{\revise{p}}_k\revise{^\mathrm{T}},~\bm{\eta}_k\revise{^\mathrm{T}},~\bm{q}_k\revise{^\mathrm{T}}]^\mathrm{T}$ and $\bm{\eta}_k=[\phi_k,~\theta_k,~\psi_k]^\mathrm{T}$ at the step $k$. 
Then, the pose of \secondRevise{the i-th} robot \secondRevise{end-effector $\bm{\xi}_{e_i}$} can be expressed by

\begin{eqnarray}
   \secondRevise{\bm{\xi}_{e_i}} = \bm{f}_{e_i}(\bm{\zeta}),
\label{eq:ee_pose}
\end{eqnarray}
When $\bm{J}_{e_i}=\partial \bm{f}/\partial \secondRevise{\bm{\zeta}} $, the relationship between \secondRevise{$\delta \bm{\xi}_{e_i}$} and $\delta \bm{\zeta}$ is expressed by
\begin{eqnarray}
   \secondRevise{\delta \bm{\xi}_{e_i}} = \bm{J}_{e_i}~\delta \bm{\zeta},
\label{eq:delta_ee}
\end{eqnarray}
where $\bm{J}_{e_i} \in \mathbb{R}^{6\times6+N_{joint}+N_{va}}$, and $N_{va}$ is the number of vectoring apparatus joints.
Here, we determine \secondRevise{$\delta \bm{\xi}_{e_i}$} as follows:

\begin{eqnarray}
   \secondRevise{\delta \bm{\xi}_{e_i}} = \kappa \dfrac{\secondRevise{{\bm{\xi}_{e_{i_d}}}} - {\bm{\xi}_{e_{i_k}}}}{||\secondRevise{{\bm{\xi}_{e_{i_d}}}} - {\bm{\xi}_{e_{i_k}}}||},
\label{eq:delta_ee_pose}
\end{eqnarray}
where ${\bm{\xi}_{e_{i_d}}}$ and ${\bm{\xi}_{e_{i_k}}}$ are the final target \secondRevise{pose} and k-th \secondRevise{pose} of the i-th \secondRevise{end-effector}, and $\kappa\add{(>0)}$ is sufficiently small value. Next, we consider the cost function $G(\bm{\zeta}_k)$.
To minimize the updated values, the first cost term is expressed as the following quadratic form:

\begin{gather}
\delta\bm{\zeta}_k^\mathsf{T} \bm{S}_1 \delta\bm{\zeta}_k.
  \label{eq:min_update_joint}
\end{gather}
According to the relationship in Eq.~\eqref{eq:delta_ee}, the constraint on the Cartesian motion in the operating space \secondRevise{$\bm{\xi}_{e_i}$} should be equality. However, it is difficult to calculate smooth trajectories using this constraint. Therefore, we consider the following quadratic residual term: 

\begin{gather}
\sum^{N_{ee}}_{i=0}(\bm{J}_{e_i}~\delta \bm{\zeta}_k - \delta \secondRevise{\bm{\xi}_{e_i}})^\mathsf{T} \bm{S}_2 (\bm{J}_{e_i}~\delta \bm{\zeta}_k - \delta \secondRevise{\bm{\xi}_{e_i}}),\label{eq:min_update_diff}
\end{gather}
where $N_{ee}$ is the number of end-effectors. $\bm{S}_1$ and $\bm{S}_2$ are the positive definite diagonal weight matrices.
For the proposed aerial manipulator, we have to consider the end-effector pose \secondRevise{$\bm{\xi}_{ee}$} and foot pose \secondRevise{$\bm{\xi}_{foot}$} as the operating space \secondRevise{$\bm{\xi}_{e_i}$}. Therefore, the cost function is given by
\begin{gather}
G(\bm{\zeta}_k)=\delta\bm{\zeta}_k^\mathrm{T} \bm{S}_1 \delta\bm{\zeta}_k + (\bm{J}_{ee}~\delta \bm{\zeta}_k - \delta \secondRevise{\bm{\xi}_{ee}})^\mathrm{T} \bm{S}_2' (\bm{J}_{ee}~\delta \bm{\zeta}_k - \delta \secondRevise{\bm{\xi}_{ee}})\nonumber\\+ (\bm{J}_{foot}~\delta \bm{\zeta}_k - \delta \secondRevise{\bm{\xi}_{foot}})^\mathrm{T} \bm{S}_2'' (\bm{J}_{foot}~\delta \bm{\zeta}_k - \delta \secondRevise{\bm{\xi}_{foot}}),
   \label{eq:opt_plan_apm}
\end{gather}
where $\bm{S}_2=\rm{diag}[\bm{S}_2',~\bm{S}_2'']^\mathrm{T}$. Furthermore, $\bm{J}_{ee}$ and $\bm{J}_{foot}$ are the \revise{Jacobians} of the end-effector and foot pose.
To generate the trajectory based on the above framework, we consider the following constraints to minimize the cost function in Eq.\eqref{eq:opt_plan}.

\subsubsection{Robot State}
The bound for \revise{the} robot state \revise{owing} to the configuration is important to generate the trajectory. Here, the velocity damper \cite{velocity_damper} is applied to avoid limitations such as joint angles. 

\begin{gather}
    \delta \bm{\zeta}^{-} \leq \delta \bm{\zeta}_k \leq \delta \bm{\zeta}^{+}, \label{eq:state_limit}
\end{gather}
\begin{eqnarray}
    \delta \bm{\zeta}^{-} \!\!&=&\!\!
        \begin{cases}
            {\delta \bm{\zeta}_{min}\dfrac{(\bm{\zeta}_k\!-\!\bm{\zeta}_{min})\!-\!\bm{\zeta}_{F}}{\bm{\zeta}_{R}-\bm{\zeta}_{F}} \ \!(\bm{\zeta}_k\!-\!\bm{\zeta}_{min} \!<\! \bm{\zeta}_{R})}\\
                {\delta \bm{\zeta}_{min} \ \rm (otherwise)}
        \end{cases}\!\!\!, \label{eq:sec5:lower_joint_torque_limit}\\
    \delta \bm{\zeta}^{+} \!\!&=&\!\!
        \begin{cases}
            {\delta \bm{\zeta}_{max}\dfrac{(\bm{\zeta}_{max}\!-\bm{\zeta}_k)\!-\!\bm{\zeta}_{F}}{\bm{\zeta}_{R}-\bm{\zeta}_{F}} \ \!(\bm{\zeta}_{max}\!-\!\bm{\zeta}_k \!<\! \bm{\zeta}_{R})}\\
                {\delta \bm{\zeta}_{max} \ \rm (otherwise)}
    \end{cases}\!\!\!, \label{eq:sec5:upper_joint_torque_limit}
\end{eqnarray}
where $\bm{\zeta}^{-}_{k}$ and $\bm{\zeta}^{+}_{k}$ are the lower and upper limitations of the i-th state, respectively. Furthermore, $\bm{\zeta}_{F}$ and $\bm{\zeta}_{R}$ denote the ranges of the restricted and forbidden state, respectively. In the restricted state, the updated values of joint angles decrease as shown in Eqs. \eqref{eq:sec5:lower_joint_torque_limit} and \eqref{eq:sec5:upper_joint_torque_limit}, and the joint angle update value becomes zero in the forbidden state.

\subsubsection{Joint Torque}
The joint torque limitation cannot be ignored while calculating the trajectory.
From the Eq.~\eqref{eq:general_model}, the joint torque in \revise{the} steady state can be expressed as follows:

\begin{align}
  \bm{\tau}_{joint} = -\bm{J}'^T_{rotor}
  \left[\begin{array}{c}
    \bm{F}_{rotor}\\
    \bm{M}_{rotor}
    \end{array}\right]
  ~~~~~~~~~~~~~~~~~~~~~~~~~~~~~~~~~~~~~ \nonumber\\
  -\sum^{N_{contact}}_{i=1}\bm{J}'^T_{contact_i}
    \left[\begin{array}{c}
    \bm{F}_{contact_i}\\
    \bm{M}_{contact_i}
    \end{array}\right]
     -\sum^{N_{part}}_{i=1}{\bm{J}'^T_{part_i}(\bm{q})
     m_{part_i}\bm{g}}.
    \label{eq:static_joint_torque}
\end{align}

Note that $\ddot{\bm{q}}$,$\revise{~\ddot{\bm{\xi}}}$, and $\bm{l}_{joint}$ are assumed to be zero\revise{, as} the maneuvering of \revise{the} joint motion of \revise{the} aerial manipulator is sufficiently slow. From the relationship in Eq.~\eqref{eq:static_joint_torque}, the joint torque constraint is given by

\begin{gather}
    \delta \bm{\tau}^{-} \leq \bm{J}_{\bm{\tau}_{i}} \delta \bm{\zeta}_k \leq \delta \bm{\tau}^{+}, \label{eq:joint_torque_limit}
\end{gather}
where the lower and upper limitations $\delta \bm{\tau}^{-}$,~$\delta \bm{\tau}^{+}$ are also expressed using the velocity damper. Furthermore, $\bm{J}_{\tau_{i}}$ is the Jacobian for the joint torque.

\subsubsection{Thrust}
For the robot equipped with rotors, the thrust limitation is critical for the maneuvering of joint motion during a flight and perching. Therefore, the constraint is given by
\begin{gather}
    \delta \bm{\lambda}^{-} \leq \bm{J}_{\lambda} \delta \bm{\zeta}_k \leq \delta \bm{\lambda}^{+}, \label{eq:rotor_force_limit}
\end{gather}
where $\delta \bm{\lambda}^{-}$ and $\delta \bm{\lambda}^{+}$ are also expressed using the velocity damper. 
Then, the upper thrust limitation during perching $\bar{\bm{\lambda}}^{+}$ is expressed as follows:

\begin{align}
    \bar{\bm{\lambda}}^{+}=\bm{\lambda}_{max}-\bm{\bar{Q}}_{perch}^{-1}(\bm{q})
    \left[\begin{matrix}
    \dfrac{1}{\mu_{xy}} \sqrt{^{}_{F}F_{ee_x}^2+^{}_{F}F_{ee_y}^2}\\
    0.0\\
    0.0\\
    0.0
    \end{matrix}\right],
\end{align}
where $\lambda_{max}$ is the thrust limitation that depends on a rotor module. $\mu_{xy}$ is the static friction coefficient. Note that $\bar{\bm{\lambda}}^{-}=0$ during perching.

\subsubsection{Stability}
The rotor arrangement is varied when the rotor-distributed robots maneuver with joint motion, and the robot cannot fly in the vicinity of a specific posture. To address this problem, we use the manipulability measure as follows: 

\begin{align}
    \rho_{stable} = \sqrt{\mathrm{det}\{\bar{\bm{Q}}(\bm{\zeta})\bar{\bm{Q}}(\bm{\zeta})^\mathrm{T}\}},
\end{align}
where robots stabilize as $\rho_{stable}$ increases.  
By differentiating the relationship, the constraints for the stability \revise{of} a rotor-distributed robot is obtained as follows:

\begin{align}
    \delta \rho_{stable} = \bm{J}_{stable}\delta \bm{\zeta}_k \geq - \gamma_{stable} \dfrac{\rho\!-\!\rho_{F}}{\rho_{R}-\rho_{F}},
    \label{eq:singurality}
\end{align}
where the velocity damper is used for this constraint. $\gamma_{stable}$ is the damping gain parameter for stability.

\subsubsection{Collision Avoidance}
Hardware interference can critically damage the robots.
To avoid collisions, we approximate the robot model using simple configurations (e.g., box, cylinder)\revise{. As} the robot is highly complex and it is difficult to consider a detailed model\revise{, the} closest point is considered. Then, the linear constraint to avoid the collision can be expressed as: 

\begin{align}
 \delta d_{collision} = \bm{n} (\bm{J}_1-\bm{J}_2)\delta \bm{\zeta}_k \geq - \gamma_{collision} \dfrac{d\!-\!d_{F}}{d_{R}-d_{F}},
   \label{eq:collision_avoidance}
\end{align}
where $\bm{n}$ is the unit vector between objects, and $\gamma_{collision}$ is the damping gain parameter used to avoid collisions. 

\subsubsection{Airflow Interference}
The airflow interference of rotors significantly affects the \revise{stability of the} robot. However, the airflow interference is considerably complex. Here, we assume the upper-stream and downstream of the rotors are \revise{cylindrical}.
Then, the distance between the i-th and j-th rotors has to satisfy the following relationship:
\begin{align}
     d_{rotor,i,j}=||\bm{p}_{rotor,i}-\bm{p}_{rotor,j}||-D_{rotor}>0,
\end{align}
where $D_{rotor}$ is the diameter of rotors. $\bm{p}_{rotor,i}$ and $\bm{p}_{rotor,j}$ are the i-th and j-th rotor positions, respectively.
From the above equation, the linear constraint on the airflow interference is expressed using the same relationship in Eq.~\eqref{eq:collision_avoidance}.

\subsubsection{Optimization Problem}
Based on the above discussion, the optimization problem in Eq.~\eqref{eq:opt_plan} is written \revise{as}
\begin{gather}
\label{eq:opt_plan_with_limit}
\min_{\delta\bm{\zeta}_k}\delta\bm{\zeta}_k^\mathrm{T} \bm{S}_1 \delta\bm{\zeta}_k + (\bm{J}_{e}~\delta \bm{\zeta}_k - \delta \secondRevise{\bm{\xi}_{ee}})^\mathrm{T} \bm{S}_2' (\bm{J}_{e}~\delta \bm{\zeta}_k - \secondRevise{\bm{\xi}_{ee}})\nonumber\\+ (\bm{J}_{foot}~\delta \bm{\zeta}_k - \delta \secondRevise{\bm{\xi}_{foot}})^\mathrm{T} \bm{S}_2'' (\bm{J}_{foot}~\delta \bm{\zeta}_k - \delta \secondRevise{\bm{\xi}_{foot}}) \label{opt_plan}
\\
   \rm{s.t.} ~~~~~~~~~~~~~~~~~~~~~~~~~~~~~~~~~~~~~~~~~~~~~~~~~~~~~~
   \\\rm{Eqs}.~\eqref{eq:state_limit},~\eqref{eq:joint_torque_limit},~\eqref{eq:rotor_force_limit},~\eqref{eq:singurality},~\rm{and}~ \eqref{eq:collision_avoidance}\nonumber.
\end{gather}
By solving this problem, we calculate the trajectories such as \revise{those of the} end-effector and achieve desired motions during perching in \revise{the} experiments \revise{described} in Sec.~\ref{sec:experiment}.

%% file: 06-system.tex
\section{System Architecture}\label{sec:system}
This section, we introduce the proposed system for the rotor-distributed manipulator. (A) First, we explain \revise{that} the control system architecture consists of the proposed flight/perching control and motion planner. (B) Next, we show the detailed configurations of the proposed RDM hardware and its inner communication system.

\subsection{Control System Architecture}\label{sec:control_system}
The \revise{entire} control system architecture is shown in Fig.~\ref{fig:control_architecture}. First, the motion planner calculates the current target states such as \revise{the} pose and joint angles based on the final target states using the differential kinematics framework as shown in Sec.\ref{sec:planning}. Based on the current target states calculated by the motion planner, the position controller determines the target attitude \revise{such as} the roll and pitch angles $\phi_d$, $\theta_d$, and vectoring apparatus angles $\alpha_i$, $\beta_i$. Next, the attitude/altitude controller calculates the rotor thrust $\bm{\lambda}_{flight}$ to achieve a stable flight. This flight controller, composed of position and attitude/altitude controllers, uses the estimated values $\bm{\tilde{\revise{p}}}$, $\bm{\tilde{\dot{\revise{p}}}}$, $\bm{\tilde{\eta}}$, \revise{and} $\bm{\tilde{\omega}}$. During perching, the QP-based controller calculates the current additional thrust to perch on ceilings $\bm{\lambda}_{perch,~t}$, using the estimated contact wrench $\bm{\tilde{W}}_{foot}$ by the foot wrench sensor. Here, the target joint and vectoring apparatus angles calculated by the motion planner and position controller are sent to servo motors, and the thrust command is sent to the rotor unit. The proposed manipulator performs motions \add{using the above system}. 

\subsection{Robot Configuration}\label{subsec:platform_experiment}
\subsubsection{Hardware}

\begin{figure}[t!]
   \begin{center}
     \includegraphics[width=0.9\columnwidth]{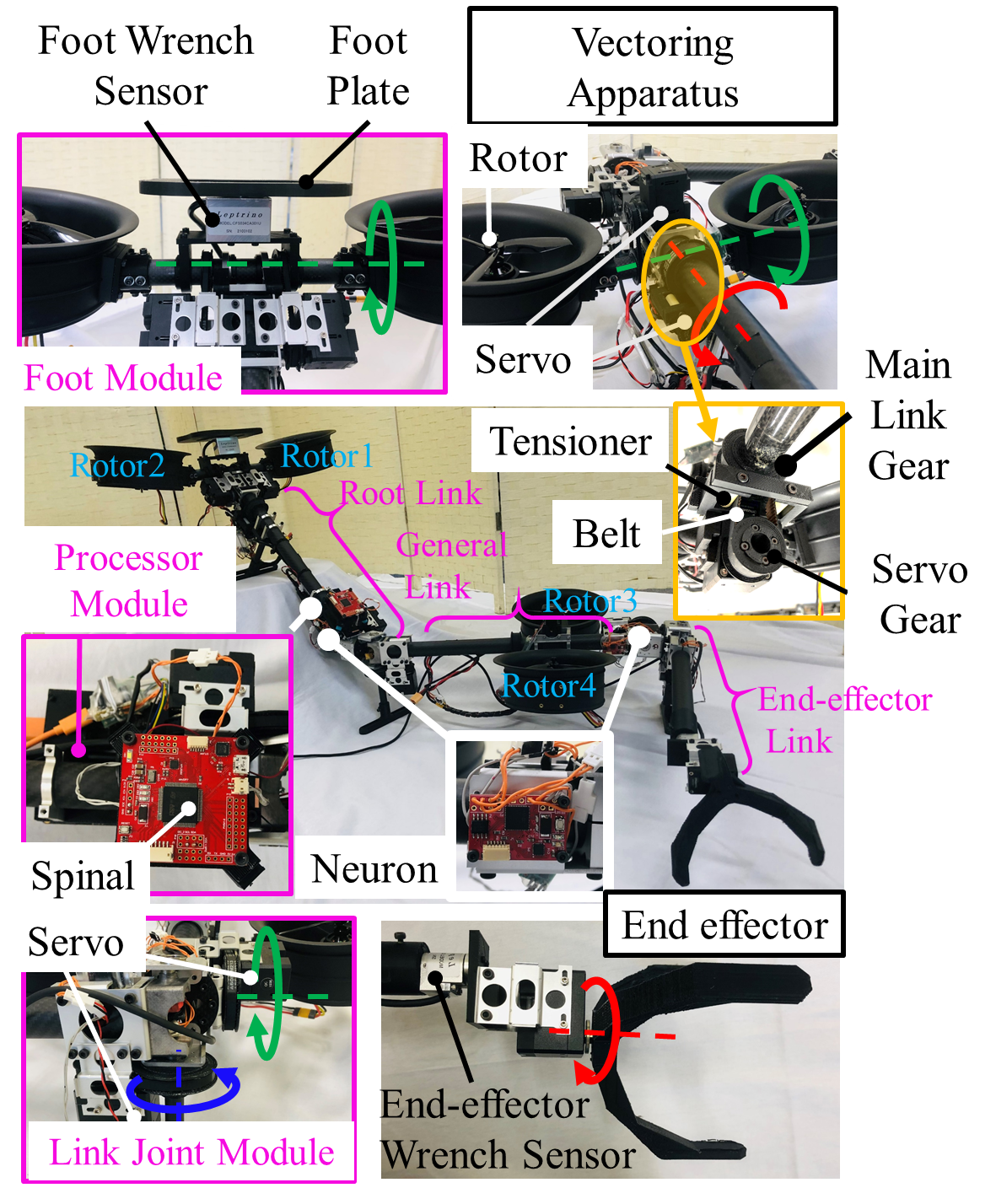}
     \caption{\revise{Proposed} platform of \revise{the} RDM.}
     \label{fig:platform}
   \end{center}
 \end{figure}

\remove{we introduce the detailed configurations of the proposed manipulator.} The main manipulator frames are made of carbon fiber reinforced plastics (CFRP) with a thickness of 1~mm, and the total mass is 3.35 kg. Furthermore, the detailed mass of each part is shown in Table~\ref{table:configuration}. Next, each link is connected by servo motors (Dynamixcel, XH430-W350-R), whose stall torque is 3.4~N \revise{in the roll and pitch directions}. The vectoring apparatuses are connected to the main links using roll (Dynamixcel, XH430-W350-R) and pitch \revise{joints} (Dynamixcel, XL430-W250-T), respectively. Here, the stall torque of the vectoring apparatus in the pitch direction is 3.4N since the roll joint moment is heavier than \revise{that of the} pitch (stall torque: 1.5N) due to the dual rotor thrust difference. Therefore, the distance between rotors on the vectoring apparatus is the smallest possible value. The rotor diameter on vectoring apparatus is $127~\mathrm{mm}$ (T-Motor) \revise{and is} actuated by a brushless motor (COBRA-2217), and can generate thrust from approximately 1-20 N. The footplate is approximated by a rectangle and equipped with a wrench sensor (CFS034CA301U, Leptrino) \revise{that} can measure the force (rated capacity:$\pm150$ N in the $x$ and $y$ directions, and $\pm300$ N in the $z$ direction) and moment (rated capacity:$\pm4.0$ Nm). Thus, the end-effector unit can be replaced by a drill, brush, or gripper. \revise{T}hese units are also equipped with a wrench sensor (CFS018CA201, Leptrino) which can measure the force (rated capacity:$\pm100$ N in the $x$ and $y$ directions, and $\pm200$ N in the $z$ direction) and moment (rated capacity:$\pm1.0$ Nm). 

\begin{table}[h!]
 \caption{\remove{The c}\add{C}onfiguration of \add{an} aerial manipulator.}
 \label{table:configuration}
 \centering
  \begin{tabular}{ccc}
   \hline
    Component & Length[m] & Mass[kg] \\ \hline\hline
    Root Link & 0.40  & 1.12\\ \hline
    General Link & 0.40 & 1.21\\ \hline
    End effector Link & 0.25 & 0.19\\ \hline
    Link Joint Module & -  & 0.13\\ \hline
    Processor Module & -  & 0.15\\ \hline
    Total Link & -   & 3.35\\ \hline
    Foot Module & -  & 0.35\\ \hline
    Foot Plate Size & 0.10$\times$0.15  \\\hline
  \end{tabular}
\end{table}

\subsubsection{Software}

 \begin{figure}[t!]
   \begin{center}
     \includegraphics[width=0.9\columnwidth]{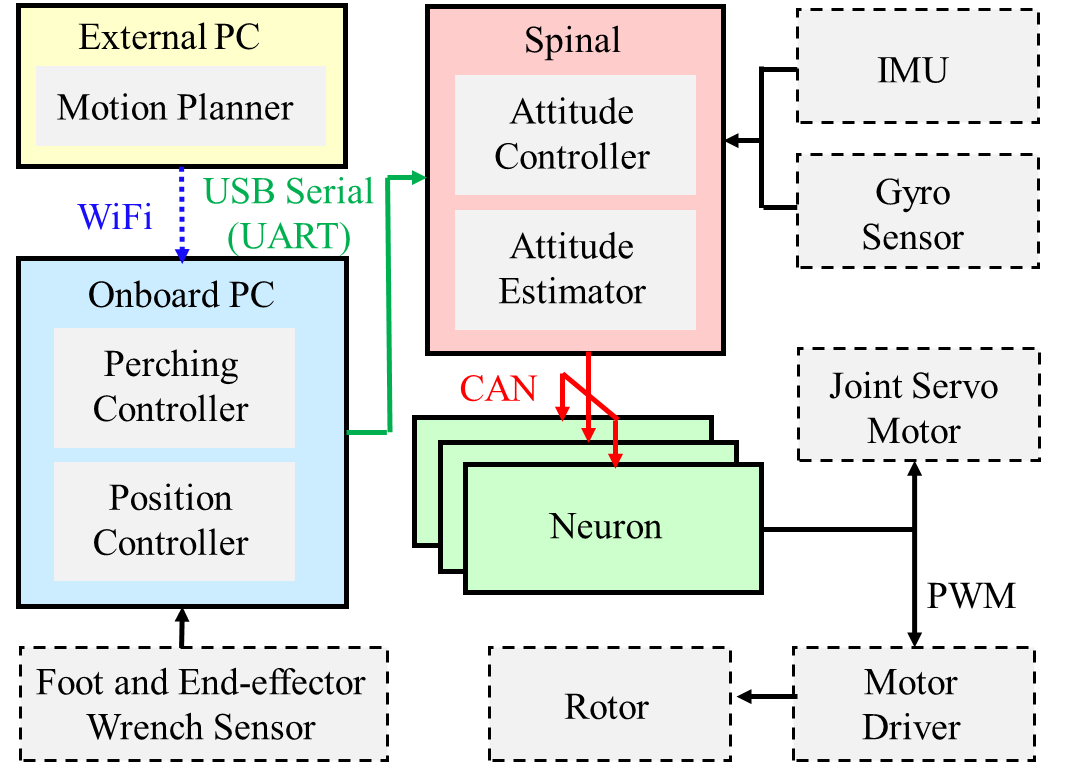}
     \caption{\remove{The s}\add{S}oftware system of the proposed RDM.}
     \label{fig:software_system}
   \end{center}
 \end{figure}

We introduce the inner communication system for the proposed manipulator as shown in Fig.~\ref{fig:software_system}. On an external PC, the motion planner calculates the target states. The data is sent to the main processor using the wireless local area network. Here, the main processor on the robot, as shown in Fig.~\ref{fig:platform} is Latte Panda Alpha 864s (Intel m3-8100Y with quad cores, clock frequency:3.4GHz). Using the data from an external PC and wrench estimated from the foot wrench sensor, the position and perching controller are running. The onboard PC communicates with the main control board (STM32F746, clock frequency:216MHz), called "Spinal" using serial communication (UART).  
On the main control board, the attitude controller and estimator are running, using the inertia moment unit (IMU) and gyro sensor, and determine the rotor thrust. Using the control area network (CAN), this main board communicates with the small control board on each link (STM32F413, clock frequency:100MHz) called "Neuron". Using RS485 and TTL signals, the servo motors on each link and vectoring apparatus are actuated by this neuron. Furthermore, the rotor unit is also actuated by the pulse width modulation (PWM) signal from the neuron. 

In this section, we introduce the control system architecture and configurations of the proposed RDM. Using this system, we conduct the experiments in Sec.~\ref{sec:experiment}.

%% file: 07-experiment.tex
\section{Experiment}\label{sec:experiment}
In this section, we evaluate the proposed RDM performance. (A) First, we evaluate the flight stability during flight. (B) Second, we confirm the perching stability (contact stability) such that the static friction and ZMP constraints are satisfied, during the maneuvering with joint motion. (C) Third, we added the external wrench at the end-effector during a flight and perching, and compared the end-effector feasibility during a flight with that \revise{during} perching. (D) Finally, several manipulation tasks were conducted during perching on actual ceilings, such as drilling, painting walls, and opening a valve. 

\subsection{Flight Stability}
\begin{figure}[t!]
   \begin{center}
     \includegraphics[width=1.0\columnwidth]{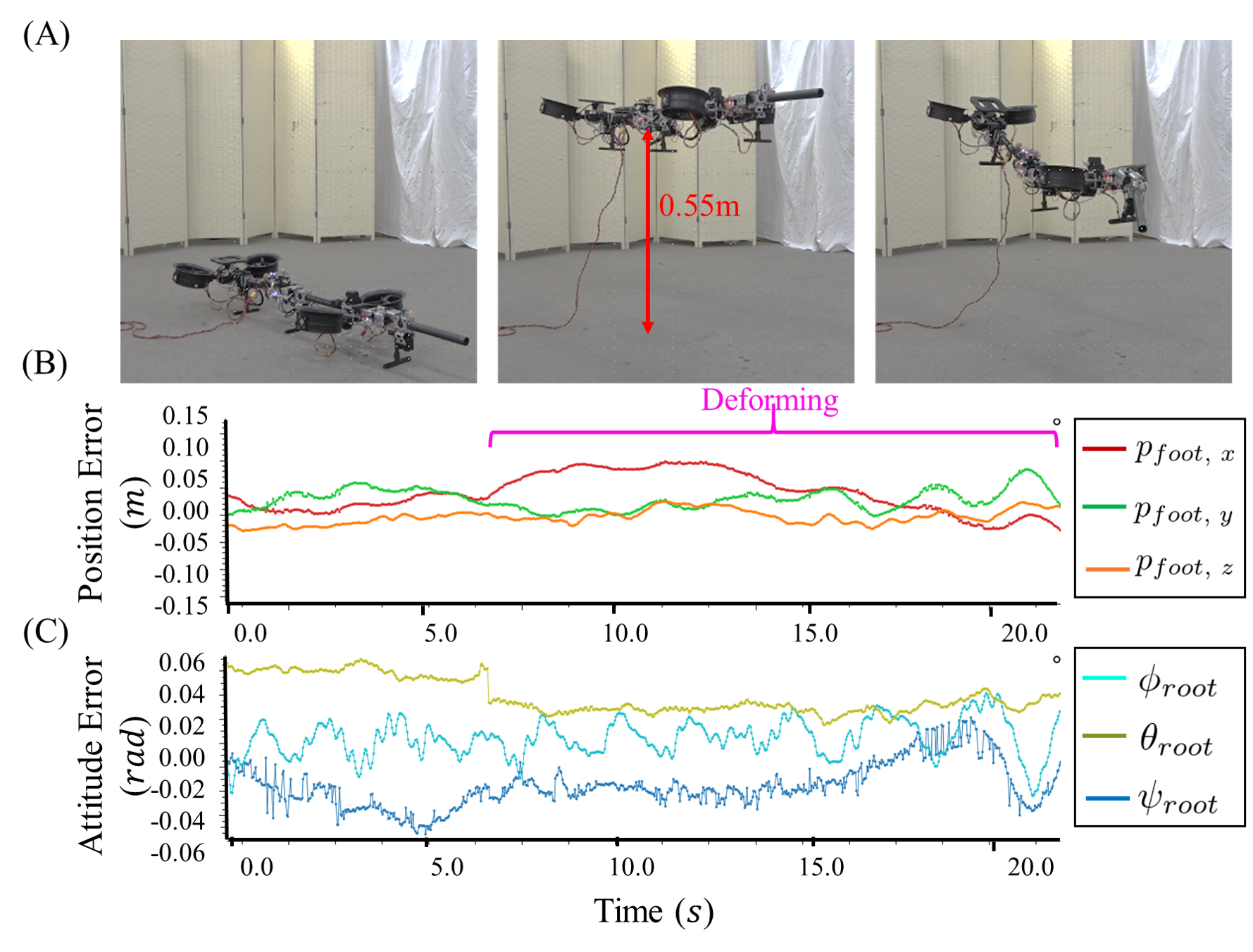}
     \caption{(A) Rotor-distributed manipulator configuration change during flight. (B) Root position tracking errors. (C) Root attitude tracking errors. }
     \label{fig:flight_test}
   \end{center}
 \end{figure}
 
We evaluated the stability during a flight.
Here, the position control rate was 40 Hz on the onboard processor, and the attitude control rate was 200Hz on the spinal.
During flight, the manipulator changed the configuration  as shown in Fig.~\ref{fig:flight_test} (A). Here, the joints between links $\bm{q}$ were varied from $[0.0,~0.0,~0.0,~0.0,~0.0,~0.0]^\mathsf{T}$ to $[0.5,~-0.5,~0.2,~0.8,~0.6,~0.0]^\mathsf{T}$. 
We set the flight control parameters as shown in Table \ref{table:flight_control_param}. The position and attitude errors became less than $\pm0.1$~m and $\pm0.05$~rad.
\begin{table}[h!]
 \caption{Flight control parameters.}
 \label{table:flight_control_param}
 \centering
  \begin{tabular}{ccc}
   \hline
   Parameter & Value & Equation\\
   \hline \hline
   $\bm{W}_1$ & $\rm {diag}[1,~1,~1,~1]$ & \multirow{2}{*}{Eq.~\eqref{eq:control_cost_func}}\\ 
   $\bm{W}_2$ & 
   \begin{tabular}{c}
   $\rm{diag}[20,~30,~90,~12.8,~600,~80$\\~~~~$,~100,~50,~20,~0.15,~1,~0.5]$
   \end{tabular} & \\\hline
   $k_{P_x},~k_{I_x}$,~$k_{D_x}$ & 4.6,~1.5,~7.0 & \multirow{2}{*}{Eq.~\eqref{eq:PID}}\\ 
   $k_{P_y},~k_{I_y}$,~$k_{D_y}$ & 7.0,~0.005,~10.5 & \\\hline 
   \add{$k_{\phi_d},k_{\theta_d}$} & 0.0, 1.0 & Eq.~\eqref{eq:attitude_angle}\\\hline 
   \remove{$u_{mask,\beta_1},u_{mask,\beta_1}$} \add{$k_{\theta_1},k_{\phi_1}$}& 0.0, 1.0 & \multirow{2}{*}{Eq.~\eqref{eq:gimbal_unit_normal}}\\
   \remove{$u_{mask,\beta_2},u_{mask,\beta_2}$} \add{$k_{\theta_2},k_{\phi_2}$} & 0.0, 0.0 & \\\hline
  \end{tabular}
\end{table}
The vectoring apparatus was designed to rotate around the main link from $-180$ to $180$ deg using a timing belt. This belt is equipped with a tensioner as shown in Fig.~\ref{fig:platform} to prevent it from being extended or slackened. The elasticity of the belt caused undesired angles of the vectoring apparatus, and the flight of aerial robots \revise{became} unstable.

The root meaning square errors (RMSEs) of the root position $\revise{\bm{p}_{foot}}$, and attitude $\revise{\bm{\eta}_{foot}}$ during the flight \revise{were} $(0.057~\rm{m},~0.033~\rm{m},~0.012~\rm{m},~0.017~\rm{rad},~0.036~\rm{rad},~0.039~\rm{rad})$ as shown in Table \ref{table:rmse_during_flight}. In this experiment, the proposed controller \revise{could} stabilize the RDM flight sufficiently.
\begin{table}[h!]
 \caption{RMSE of the proposed manipulator during flight.}
 \label{table:rmse_during_flight}
 \centering
  \begin{tabular}{cll}
   \hline
   {} & Position (m) & Attitude (rad)\\
   \hline \hline
   $x$ & 0.057  & 0.017\\ 
   $y$ & 0.033 & 0.036\\ 
   $z$ & 0.012 & 0.039\\
   \hline
  \end{tabular}
\end{table}

\subsection{Perching Stability}
\begin{figure}[t!]
   \begin{center}
     \includegraphics[width=1.0\columnwidth]{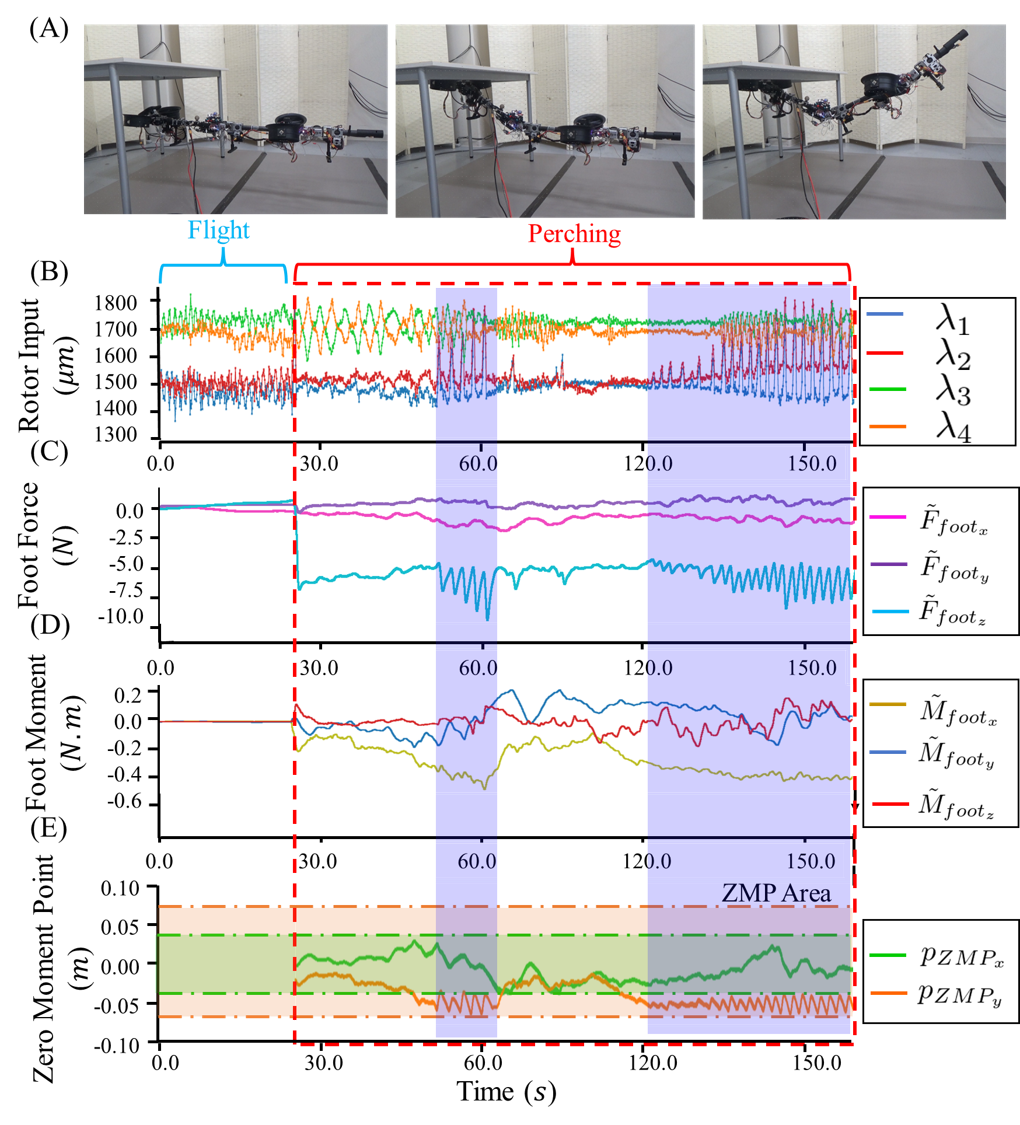}
     \caption{(A) Transition from flight to perching and \remove{deformation} configuration change during perching. (B) Rotor inputs (C) Foot force observed by a wrench sensor. (D) Foot moment observed by a wrench sensor. (E) Foot ZMP. }
     \label{fig:perching_stability}
   \end{center}
 \end{figure}

We evaluated the manipulator stability during the perching motion. The proposed perching controller calculates the additional thrust using an onboard processor at the rate of 40Hz. Here, we used the OSQP solver to optimize the QP problem in Eq.~\eqref{eq:qp}.

\begin{table}[b!]
 \caption{\remove{The }Perching \remove{C}\add{c}ontrol \remove{P}\add{p}arameters.}
 \label{table:perching_control_param}
 \centering
  \begin{tabular}{ccc}
   \hline
   Parameter & Value & Equation\\
   \hline \hline
   $\bm{N}$ & $\rm{diag}[1.0,~1.0,~1.0,~1.0]$ & Eq.\eqref{eq:qp_opt_perch}\\\hline
   $\bm{U}_{mask}$ & $\rm{diag}[1.0,~0.0,~0.0,~0.0]$ &  \multirow{2}{*}{Eq.~\eqref{eq:perching_thrust}}\\ 
   $\bm{\bar{Q}}_{perch}$ & $\bm{\bar{Q}}(\bm{q})\rm{diag}[1, 1, 0, 0]$ & \\\hline
   $\epsilon_z$ & 1.0 & Eq.~\eqref{eq:floor_reaction}\\ \hline
   $\mu_x,~\mu_y$ & 0.9 & Eq.~\eqref{eq:sliding_prevention}\\\hline 
   $\mu_z$ & 0.5 & Eq.~\eqref{eq:rotating_prevention}\\\hline
   $H_x,~H_y$ & 0.04, 0.07 &  Eqs.\eqref{eq:zmp_constraint_reqtangular}\\\hline
   $\delta\bm{\lambda}_{perch}$ & $[5.0,~5.0,~5.0,~5.0]^\mathsf{T}$ & Eq.~\eqref{eq:continuous_thrust}\\\hline
  \end{tabular}
\end{table}

We set the perching control parameters to prevent slipping and peeling from ceilings, as shown in Table~\ref{table:perching_control_param}. Here, the estimated force and torque by the wrench sensor shows significant amount of noise at times. The controller \revise{is required to generate an} excessive thrust for perching. In this situation, the target rotor thrust exceeds the upper thrust limitation, and the manipulator can become unstable. Therefore, the infinite impulse response (IIR) filter, whose cutoff frequency is 1Hz, was used to smoothen the data.

We experimentally achieved a stable maneuvering with the joint motion of the proposed manipulator during perching, as shown in Fig.~\ref{fig:perching_stability} (A). 
Here, the manipulator approached the ceiling and could smoothly shift from flight mode to perching mode at 27s. 
During the transition from a flight to perching, the rotor inputs were significantly increased to suppress the unstable contact such as slipping and peeling from the ceilings. The thrust of the foot rotors became approximately 1.4 times larger than that during the flight for a moment, as shown in Fig.~\ref{fig:perching_stability} (B). However, the manipulator became stable after that, and the foot rotor thrust \revise{decreased} and becomes as large as that of the flight. This result shows that the thrust and consumption power for stable perching decreases \revise{owing} to the ceiling effect. However, the rotor thrust fluctuated from 52s to 62s and from 135s to 160s when $p_{ZMP_x}$ approached the boundaries during maneuvering with joint motion. This is because the controller had to increase the thrust to prevent the footplate from peeling from the ceilings. In this situation, power consumption also increased.
During perching, the foot force $\tilde{F}_{foot_x},~\tilde{F}_{foot_y},~\tilde{F}_{foot_z}$ and the moment $\tilde{M}_{foot_z}$ satisfied the static friction constraints in Eqs. \eqref{eq:floor_reaction}, \eqref{eq:sliding_prevention}, and \eqref{eq:rotating_prevention} with a margin, as shown in Figs.~\ref{fig:perching_stability} (C) and (D). Furthermore, the ZMP \revise{was} within the footplate area, as shown in Fig.~\ref{fig:perching_stability} (E). These results \revise{reveal} that the footplate can perch on the ceiling with a significant amount of stability.
Here, the rotor thrust during perching is equal to the thrust during flight, although the $\tilde{F}_{foot_z}$ increased to fix the footplate on the ceiling.

\subsection{End-Effector Evaluations}
For the manipulation tasks, it is critical to evaluate the end-effector performance. Here, we conducted two experiments: (1) \revise{we} added the external wrench at the end effector to evaluate the feasible wrench, and (2) compared the end-effector \revise{pose} tracking stability during a flight with perching.  

\subsubsection{External Wrench at the End-effector}
\begin{table}[b!]
 \caption{\add{Feasible force at the end-effector.}}
 \label{table:ee_wrench}
 \centering
  \begin{tabular}{cllll}
   \hline
   {} & \multicolumn{2}{c}{\add{Flight}} & \multicolumn{2}{c}{\add{Perching}} \\
   {} & \add{Average} & \add{Peak} & \add{Average} & \add{Peak}\\
   \hline \hline
   \add{$|F_{ee,~x}|$} & \add{3.75}  & \add{7.48}& \add{25.1}  & \add{41.8}\\ 
   \add{$|F_{ee,~y}|$} & \add{1.76}  & \add{2.62} & \add{8.4} & \add{15.9}\\ 
   \add{$|F_{ee,~z}|$} & \add{3.42}  & \add{6.07} & \add{16.8} & \add{32.1}\\
   \add{$|M_{ee,~x}|$} & \add{0.054}  & \add{0.082} & \add{0.736}  & \add{1.266}\\ 
   \add{$|M_{ee,~y}|$} & \add{0.620}  & \add{0.975} & \add{0.298} & \add{0.560}\\ 
   \add{$|M_{ee,~z}|$} & \add{0.0203}  & \add{0.197} & \add{0.943} & \add{1.520}\\
   \hline
  \end{tabular}
\end{table}

\begin{figure}[t!]
   \begin{center}
     \includegraphics[width=1.0\columnwidth]{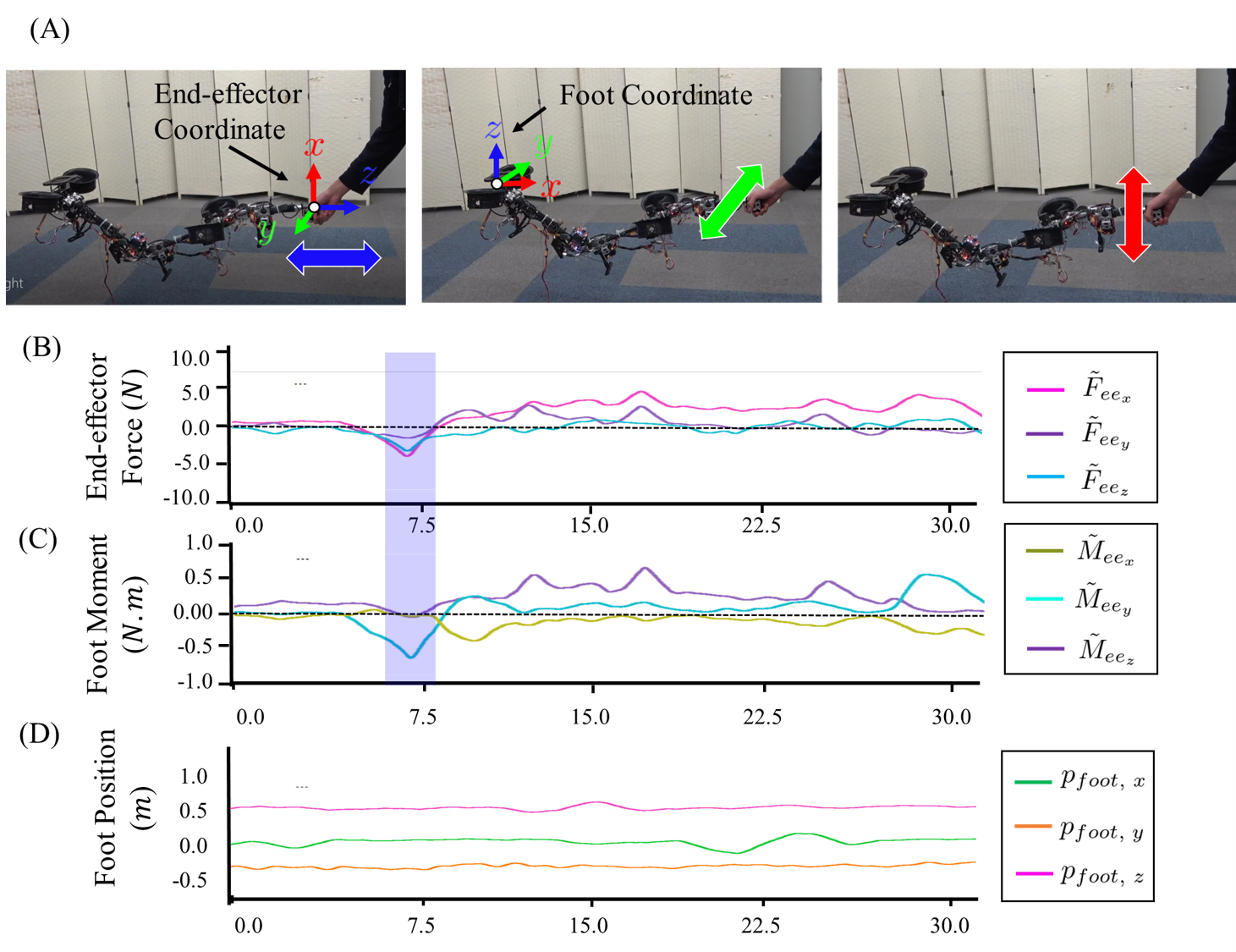}
     \caption{Manipulator flight stability under end-effector wrench disturbances. (A) The end-effector is pushed and pulled; (B) end-effector force observed by a wrench sensor; (C) end-effector moment observed by a wrench sensor; and (D) foot position observed by a motion capture system (ground truth). }
     \label{fig:ee_wrench_flight}
   \end{center}
 \end{figure}

\begin{figure}[b!]
   \begin{center}
     \includegraphics[width=1.0\columnwidth]{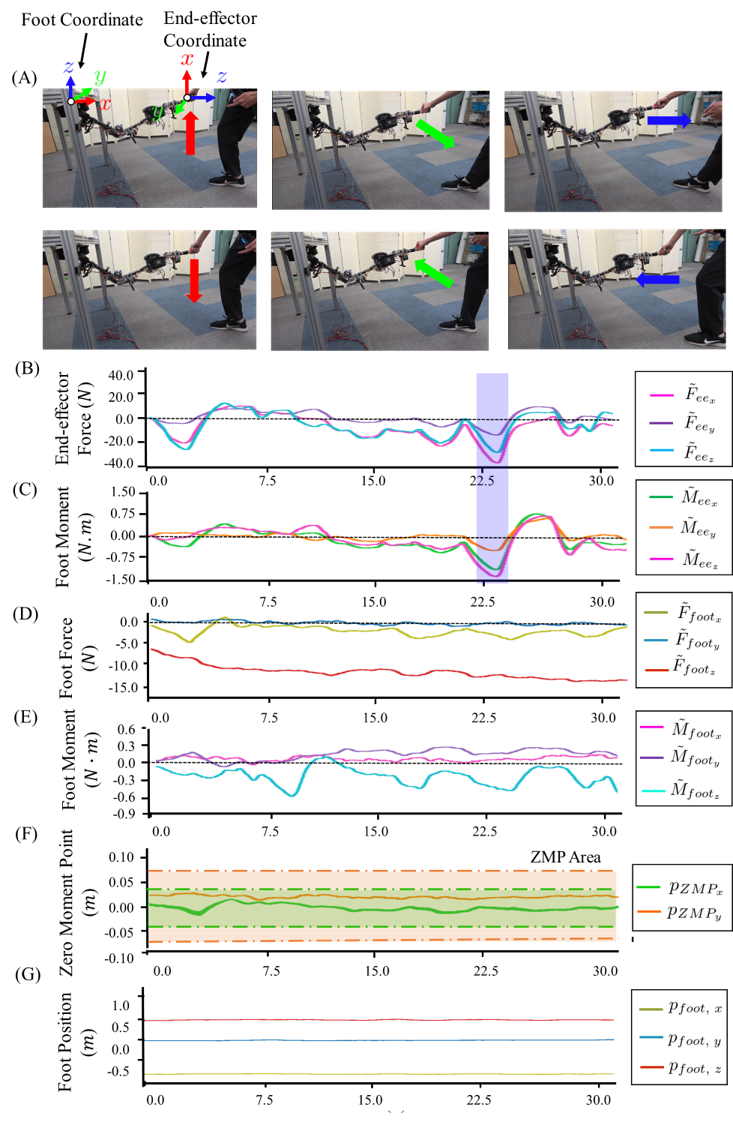}
     \caption{Manipulator stability under end-effector wrench disturbances. (A) We push and pull the end-effector. (B) End-effector force observed by a wrench sensor. (C) End-effector moment observed by a wrench sensor. (D) Rotor inputs (E) Foot force observed by a wrench sensor. (F) Foot moment observed by a wrench sensor. (G) Foot ZMP. (H) Foot position observed by the motion capture system (ground truth)}
     \label{fig:ee_wrench}
   \end{center}
 \end{figure}

\begin{table}[h!]
 \caption{\add{Standard deviation of the foot position during flight and perching.}}
 \label{table:standard_deviation_foot}
 \centering
  \begin{tabular}{cll}
   \hline
   {} & \add{Flight} & \add{Perching} \\
   \hline \hline
   \add{$x$} & \add{0.0416}  & \add{0.0009}\\ 
   \add{$y$} & \add{0.0281} & \add{0.0011}\\ 
   \add{$z$} & \add{0.0220} & \add{0.0004}\\
   \hline
  \end{tabular}
\end{table}

For manipulation tasks, it is critical to evaluate stability and feasible wrench when the external wrench is added to the end-effector. Therefore, we pushed and pulled the end-effector in the $x$, $y$, and $z$ directions during the flight and perching without the external wrench observation control as shown in Figs.\ref{fig:ee_wrench_flight} and \ref{fig:ee_wrench}. 

In the flight experiment, the added wrench in the end-effector coordinate is shown in Figs.\ref{fig:ee_wrench_flight} (B) and (C). Here, we focused on the time from 5.5 to 8.0s. The absolute values of the average end-effector force $F_{ee,~x}$, $F_{ee,~y}$, and $F_{ee,~z}$ were 3.75, 1.76, and 3.42~N. As shown in Fig.\ref{fig:ee_wrench_flight} (D), the foot position was varied under external wrench disturbances, and the standard deviation of the foot position in \revise{the} $x$, $y$, and $z$ directions were 0.0416, 0.0281, and 0.0220~m\revise{, respectively,} as shown in Table.\ref{table:standard_deviation_foot}.

In the perching experiment, the added end-effector wrench is shown in Figs.\ref{fig:ee_wrench} (B) and (C). 
Here, the time from 22 to 24s was considered. The absolute values of the average forces during this time were 25.1, 8.4, and 16.8~N. In this configuration, the feasible force $F_{ee,y}$ was comparatively smaller than $F_{ee,x}$ and $F_{ee,z}$ \revise{owing} to the yaw control feasibility. This is because a small $F_{ee,y}$ generates a large moment in the yaw direction at the origin of the foot coordinate. Therefore, the static friction moment constraint could not be satisfied, and the footplate slips in the yaw direction. \revise{Owing} to the external disturbance, the foot force $F_{foot, z}$ in the foot coordinate increases significantly as shown in Fig.\ref{fig:ee_wrench} (D) to prevent the foot from slipping and peeling from the ceilings. Owing to this thrust increase, the static friction and ZMP constraints were satisfied as shown in Fig\revise{s}.\ref{fig:ee_wrench} (D), (E), and (F). 
Furthermore, the standard deviation of the foot position during perching in the $x$-, $y$-, and $z$- directions \revise{were} 0.0009, 0.0011, and 0.0004m, respectively\revise{. They were} significantly smaller than that during flight as shown in Fig.\ref{fig:ee_wrench} (F) and Table \ref{table:ee_wrench}. This phenomenon was observed even \revise{when} the added external force during perching was larger than that during the flight.
These experiments clarified that the manipulator generates \revise{a} larger wrench at the end-effector by \revise{being} fixed on ceilings.

  \begin{figure*}[th!]
   \begin{center}
     \includegraphics[width=2.0\columnwidth]{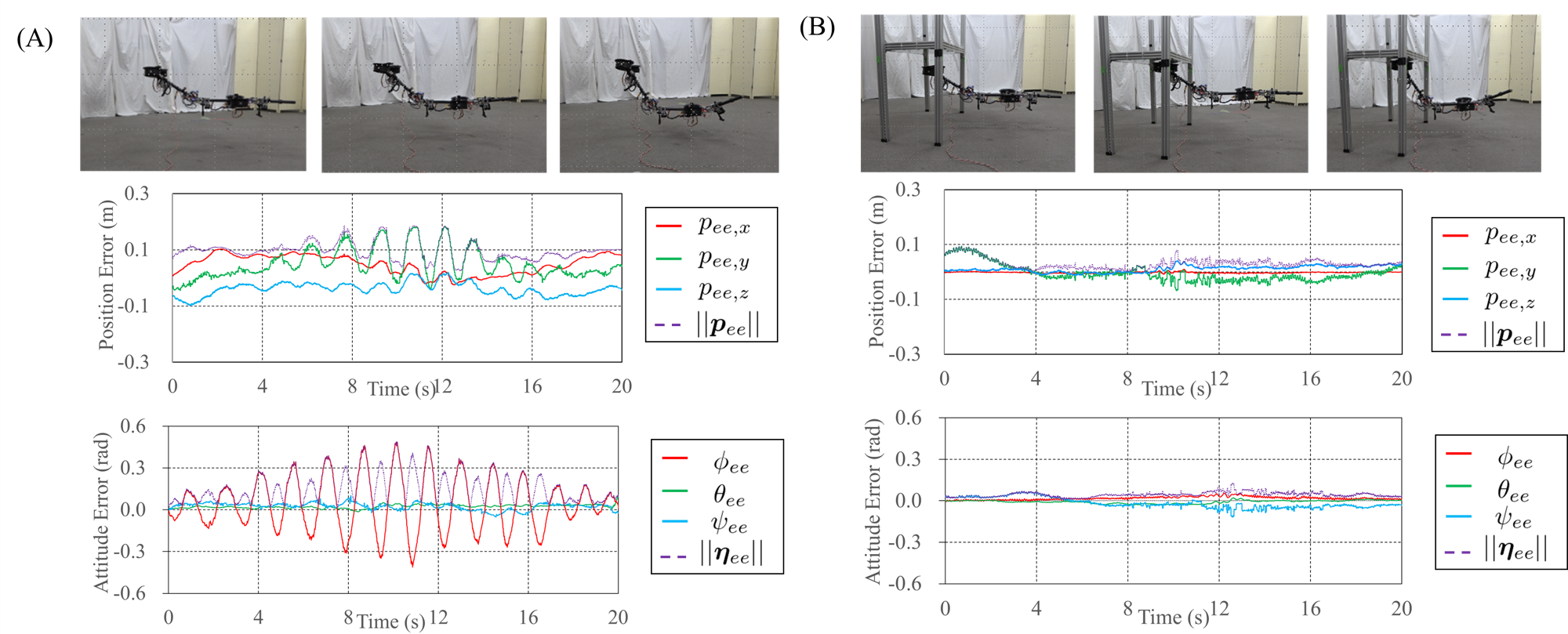}
     \caption{\remove{The c}\add{C}omparison \remove{at}\add{of} the end-effector \revise{position and attitude} of \add{the proposed} rotor-distributed \remove{aerial} manipulator. (A) \remove{D}\add{d}uring flight (B) \remove{D}\add{d}uring perching. \add{In these experiments, the baselink position and attitude are controlled by the real-time feedback method as shown in Sec.~\ref{sec:control}. However, the joint angles are determined by a feed-forward motion planner.}}
     \label{fig:compare_stability}
   \end{center}
 \end{figure*}

 \subsubsection{Stability at the End-Effector}
We developed \revise{an} RDM with a perching ability to improve the end-effector stability.
To evaluate the end-effector stability during flight and perching, we compared the position \revise{and attitude} tracking ability at the end-effector. Here, \revise{$\bm{p}_{ee}$ and $\bm{\eta}_{ee}$} were the \revise{position and attitude} at the end effector, respectively.
The target \secondRevise{pose} is determined by the planner. Note that we used feedback control not in the motion planner but in position \secondRevise{and attitude} control.
As shown in Fig.~\ref{fig:compare_stability} (A), the end-effector during flight became unstable, especially from \revise{5}\remove{s} to \revise{16} s \revise{owing} to disturbances such as the ground effect, and the \revise{maximum amplitudes of position and attitude errors were over 0.1 m and 0.3 rad, respectively.} It is difficult for aerial robots, especially during maneuvering with joint motion, to stabilize the position under disturbances \revise{owing} to the complexity of the rotor airflow dynamics. However, the end-effector during perching became significantly stable, and the end-effector did not vibrate. \revise{Owing} to the fixed root approach, the end-effector position \revise{and attitude} errors were \revise{generally} less than 0.05~m and \revise{0.1 rad} as shown in Fig.~\ref{fig:compare_stability} (B). Furthermore, the \revise{position and attitude} RMSE during flight and perching \revise{were} $[0.055~\rm{m},~0.076~\rm{m},~0.069~\rm{m}$, $\revise{0.0019~\rm{rad},~0.0003~\rm{rad},~0.0004~\rm{rad}}]$ and $[0.0032~\rm{m}$, $0.029~\rm{m},~0.019~\rm{m}$, $\revise{0.0002~\rm{rad},~0.0001~\rm{rad},~0.0005~\rm{rad}}]$ as shown in Table. \ref{table:stability}. These results clarified that the end-effector position tracking performance significantly improved due to the fixed-root approach. 

 \begin{table}[h!]
 \caption{\remove{The }RMSE at the end-effector during flight and perching.}
 \label{table:stability}
 \centering
  \begin{tabular}{cll}
   \hline
   {} & Flight & Perching \\
   \hline \hline
   $x$ & 0.0552  & 0.00320\\ 
   $y$ & 0.0759 & 0.0286\\ 
   $z$ & 0.0688 & 0.0191\\
   \revise{$\phi$} & \revise{0.0019}  & \revise{0.0002}\\
   \revise{$\theta$} & \revise{0.0003} & \revise{0.0001}\\ 
   \revise{$\psi$} & \revise{0.0004} & \revise{0.0005}\\
   \hline
  \end{tabular}
\end{table}

\subsection{Perching Manipulation Tasks}
\begin{figure*}[th!]
   \begin{center}
     \includegraphics[width=1.95\columnwidth]{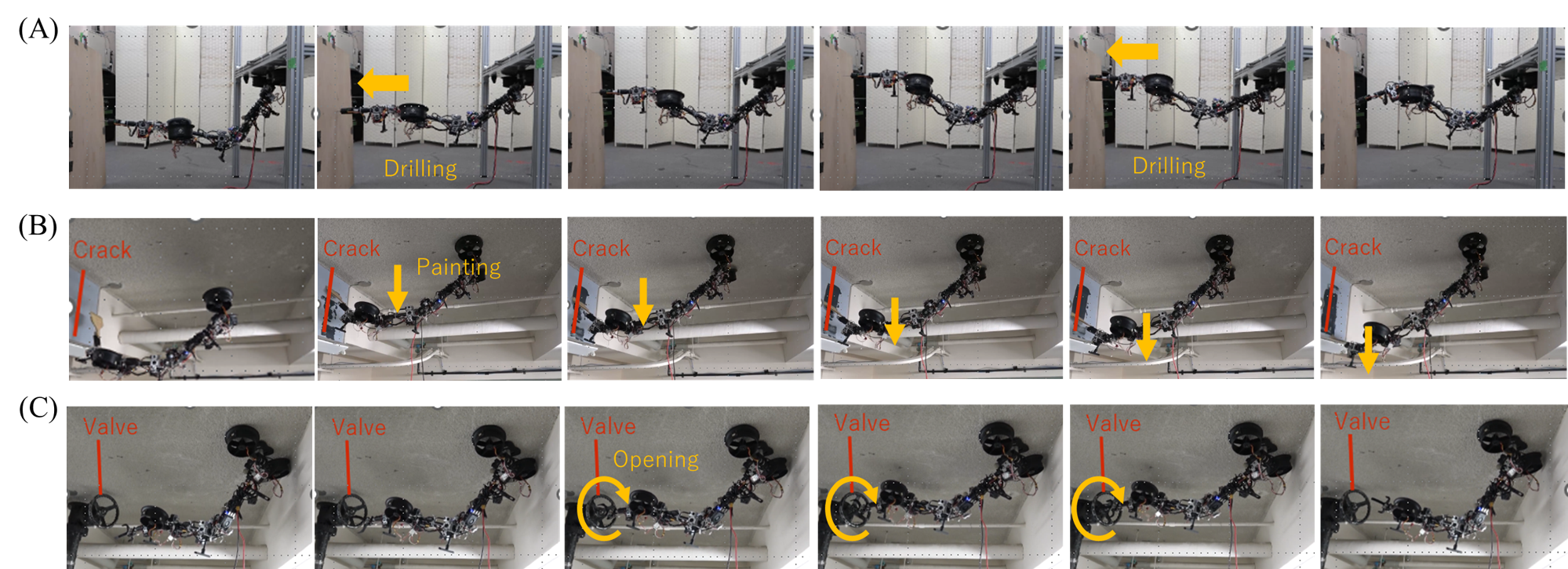}
     \caption{Manipulation tasks by rotor-distributed aerial manipulator on the ceilings: (A) drill sampling, (B) painting walls and (C) opening a valve.}
     \label{fig:task}
   \end{center}
 \end{figure*}

Using the proposed control and planning method for the rotor-distributed manipulator, we achieved manipulation on the ceiling as shown in Fig.~\ref{fig:task}. In Fig.~\ref{fig:task} (A), the manipulator conducted a drill manipulation. This manipulation task is classified as the point-contact task, and the end-effector can generate sufficient force to drill several holes in a wood board, and the position errors of these holes became approximately 0.01m. To increase the precision at the end-effector, real-time position feedback control would be one of the most useful approaches.
In addition, the manipulator perches on the actual ceiling and paints walls using a brush, as shown in Fig.\ref{fig:task} (B). This task is classified as the sliding task, and the sliding area length was over 0.5m in the vertical direction. Finally, this manipulator conducted a more complex manipulation task. This robot generated the moment at the end-effector gripper and opened a valve, as shown in Fig.\ref{fig:task} (C). These experiments \revise{revealed} the versatility of the proposed manipulator.
In these experiments, the power of the robot was supplied by a cable. 

%% file: 08-conclusion.tex
\section{Conclusion}\label{sec:conclusion}
In this study, we developed a minimal RDM configuration for perching. Based on the proposed system, we demonstrated that the RDM achieved stable maneuvering with joint motion during flight/perching and performed several manipulation tasks. The following are \revise{the} three key contributions of our approach:
(1) Considering the aerial ZMP and end-effector reachability, we proposed a quadrotor-based RDM as the minimal configuration \revise{for an RDM} and demonstrated that the \revise{realized end-effector feasible wrench and reachability of the RDM were significantly better than those of the conventional RCMs}. (2) Using the proposed QP-based controller considering the foot contact conditions, the end-effector became more precise and generated a larger wrench during perching. Furthermore, this control method decreased the desired thrust and power consumption for perching using the ceiling effect. This controller can \revise{also} be adopted \revise{for} conventional aerial robots. (3) The proposed RDM achieved several manipulation tasks on actual ceilings using the proposed motion planning method considering constraints such as \revise{the} perching force. 

In \revise{a} future work, we \revise{could} further improve the precision at the end-effector and decrease the desired thrust and power consumption during perching. For more precise manipulation, we would develop a real-time feedback controller for the end-effector position based on the proposed motion planner. To decrease the thrust and power consumption, \revise{the thrust in the horizontal directions \revise{would} be utilized to compensate for the external disturbances at the end-effector.}
\revise{In the proposed controller, only vertical force is used to perch on surfaces, as this approach is comparatively easy to stabilize the robot during perching. However, this controller requires a significant perching force when the external disturbance is added at the end-effector. This is because the feasible static friction to prevent slipping is comparatively small. To increase energy efficiency and generate a larger wrench at the end-effector, a control approach which uses both horizontal-direction thrust and static friction is required.}  
Furthermore, we could extend the perchable environment, such as side walls or pipes, and achieve dual-arm manipulation for more complex manipulation tasks.
We believe that our proposed system \revise{could} extend the applications of aerial manipulators, such as structural inspections and repairs.

%% file: appendix.tex
\appendices

\section{Flight Stability of Rotor Distributed Robots}\label{appendix:aerial_stability}
The rotor force and moment at the $\bm{p}_{AZMP}$ can be expressed as
\begin{align}
    \bm{F}_{rotor,AZMP} =\sum_{i=1}^{N_{rotor}}\bm{F}_{rotor,i},~~~~~~~\\
    \bm{M}_{rotor,AZMP} = \sum_{i=1}^{N_{rotor}}(\bm{p}_{rotor,i}-\bm{p}_{AZMP}) \times \bm{F}_{rotor,i},
    \label{eq:feasible_torque}
\end{align}
where the stable control can be achieved where the moments caused by rotors in the roll and pitch directions are zeros. Then, $\bm{p}_{AZMP_{xy}}$ is given by
\begin{align}
    \bm{p}_{AZMP_{xy}} = \dfrac{\sum_{i=1}^{N_{rotor}}\bm{p}_{rotor,i_{xy}}F_{rotor,i_z}}{\sum_{i=1}^{N_{rotor}}F_{rotor,i_z}}.
\end{align}

Here, $F_{rotor,i_z} \geq 0$, and we define $\gamma_i = \dfrac{F_{rotor,i_z}}{\sum_{i=1}^{N_{rotor}}F_{rotor,i_z}}$.
Then, the $\gamma_i$ can be as follows:
\begin{align}
    \gamma_i \geq 0, \\
    \sum_{i=1}^{N_{rotor}}\gamma_i=1 .\label{eq:gamma_limitation}
\end{align}

Based on the above discussion, $\bm{p}_{AZMP_{xy}}$ constraints to achieve a stable flight can be written as follows:
\begin{align}
    \bm{p}_{AZMP_{xy}} \in \sum_{i=1}^{N_{rotor}}\gamma_i\bm{p}_{rotor,i_{xy}}
    \label{eq:stable_boundary}
\end{align}

We call this area the aerial support polygon and use it to evaluate the flight stability in Sec.~\ref{sec:design}.

%% file: bare_jrnl.bbl
\begin{thebibliography}{10}

\bibitem{aerial_manipulator_general}
Anibal Ollero, Marco Tognon, Alejandro Suarez, Dongjun Lee, and Antonio
  Franchi.
\newblock Past, present, and future of aerial robotic manipulators.
\newblock {\em IEEE Transactions on Robotics}, 38(1):626--645, 2021.

\bibitem{helicopter}
K.~Kondak, F.~Huber, M.~Schwarzbach, M.~Laiacker, D.~Sommer, M.~Bejar, and
  A.~Ollero.
\newblock Aerial manipulation robot composed of an autonomous helicopter and a
  7 degrees of freedom industrial manipulator.
\newblock In {\em 2014 IEEE International Conference on Robotics and Automation
  (ICRA)}, pages 2107--2112, 2014.

\bibitem{rigid-arm}
H.~W. Wopereis, J.~J. Hoekstra, T.~H. Post, G.~A. Folkertsma, S.~Stramigioli,
  and M.~Fumagalli.
\newblock Application of substantial and sustained force to vertical surfaces
  using a quadrotor.
\newblock In {\em 2017 IEEE International Conference on Robotics and Automation
  (ICRA)}, pages 2704--2709, 2017.

\bibitem{rigid-arm2}
Miguel~{\'A}ngel Trujillo, Jos{\'e}~Ramiro Mart{\'\i}nez-de Dios, Carlos
  Mart{\'\i}n, Antidio Viguria, and An{\'\i}bal Ollero.
\newblock Novel aerial manipulator for accurate and robust industrial ndt
  contact inspection: A new tool for the oil and gas inspection industry.
\newblock {\em Sensors}, 19(6):1305, 2019.

\bibitem{rigid-arm3}
Karen Bodie, Maximilian Brunner, Michael Pantic, Stefan Walser, Patrick
  Pf{\"a}ndler, Ueli Angst, Roland Siegwart, and Juan Nieto.
\newblock An omnidirectional aerial manipulation platform for contact-based
  inspection.
\newblock {\em arXiv preprint arXiv:1905.03502}, 2019.

\bibitem{rigid-arm4}
Salua Hamaza, Ioannis Georgilas, Manuel Fernandez, Pedro Sanchez, Thomas
  Richardson, Guillermo Heredia, and Anibal Ollero.
\newblock Sensor installation and retrieval operations using an unmanned aerial
  manipulator.
\newblock {\em IEEE Robotics and Automation Letters}, 4(3):2793--2800, 2019.

\bibitem{rigid-arm5}
T.~Bartelds, A.~Capra, S.~Hamaza, S.~Stramigioli, and M.~Fumagalli.
\newblock Compliant aerial manipulators: Toward a new generation of aerial
  robotic workers.
\newblock {\em IEEE Robotics and Automation Letters}, 1(1):477--483, 2016.

\bibitem{rigid-arm6}
Burak Yüksel, Saber Mahboubi, Cristian Secchi, Heinrich~H. Bülthoff, and
  Antonio Franchi.
\newblock Design, identification and experimental testing of a light-weight
  flexible-joint arm for aerial physical interaction.
\newblock In {\em 2015 IEEE International Conference on Robotics and Automation
  (ICRA)}, pages 870--876, 2015.

\bibitem{rigid-arm-6d}
Markus Ryll, Giuseppe Muscio, Francesco Pierri, Elisabetta Cataldi, Gianluca
  Antonelli, Fabrizio Caccavale, Davide Bicego, and Antonio Franchi.
\newblock 6d interaction control with aerial robots: The flying end-effector
  paradigm.
\newblock {\em The International Journal of Robotics Research},
  38(9):1045--1062, 2019.

\bibitem{rigid-arm-6d-2}
Karen Bodie, Maximilian Brunner, Michael Pantic, Stefan Walser, Patrick
  Pfändler, Ueli Angst, Roland Siegwart, and Juan Nieto.
\newblock Active interaction force control for contact-based inspection with a
  fully actuated aerial vehicle.
\newblock {\em IEEE Transactions on Robotics}, 37(3):709--722, 2021.

\bibitem{rigid-arm-6d-3}
Ramy Rashad, Davide Bicego, Jelle Zult, Santiago Sanchez-Escalonilla, Ran Jiao,
  Antonio Franchi, and Stefano Stramigioli.
\newblock Energy aware impedance control of a flying end-effector in the
  port-hamiltonian framework.
\newblock {\em IEEE Transactions on Robotics}, 38(6):3936--3955, 2022.

\bibitem{5dof-arm}
Carmine~Dario Bellicoso, Luca~Rosario Buonocore, Vincenzo Lippiello, and Bruno
  Siciliano.
\newblock Design, modeling and control of a 5-dof light-weight robot arm for
  aerial manipulation.
\newblock In {\em 2015 23rd Mediterranean Conference on Control and Automation
  (MED)}, pages 853--858, 2015.

\bibitem{dual-arm2}
Alejandro Suarez, Fran Real, Víctor~M. Vega, Guillermo Heredia, Angel
  Rodriguez-Castaño, and Anibal Ollero.
\newblock Compliant bimanual aerial manipulation: Standard and long reach
  configurations.
\newblock {\em IEEE Access}, 8:88844--88865, 2020.

\bibitem{tilt-3dof}
Gabriele Nava, Quentin Sablé, Marco Tognon, Daniele Pucci, and Antonio
  Franchi.
\newblock Direct force feedback control and online multi-task optimization for
  aerial manipulators.
\newblock {\em IEEE Robotics and Automation Letters}, 5(2):331--338, 2020.

\bibitem{6dof-arm}
R~Cano, C~P{\'e}rez, F~Pruano, A~Ollero, and G~Heredia.
\newblock Mechanical design of a 6-dof aerial manipulator for assembling bar
  structures using uavs.
\newblock In {\em 2nd RED-UAS 2013 workshop on research, education and
  development of unmanned aerial systems}, volume 218, 2013.

\bibitem{7dof-arm}
G.~Heredia, A.E. Jimenez-Cano, I.~Sanchez, D.~Llorente, V.~Vega, J.~Braga, J.A.
  Acosta, and A.~Ollero.
\newblock Control of a multirotor outdoor aerial manipulator.
\newblock In {\em 2014 IEEE/RSJ International Conference on Intelligent Robots
  and Systems}, pages 3417--3422, 2014.

\bibitem{dual-arm}
Matko Orsag, Christopher Korpela, Stjepan Bogdan, and Paul Oh.
\newblock Dexterous aerial robots—mobile manipulation using unmanned aerial
  systems.
\newblock {\em IEEE Transactions on Robotics}, 33(6):1453--1466, 2017.

\bibitem{triple-arm}
Hannibal Paul, Ryo Miyazaki, Robert Ladig, and Kazuhiro Shimonomura.
\newblock Landing of a multirotor aerial vehicle on an uneven surface using
  multiple on-board manipulators.
\newblock In {\em 2019 IEEE/RSJ International Conference on Intelligent Robots
  and Systems (IROS)}, pages 1926--1933, 2019.

\bibitem{actuation-definition}
Russ Tedrake.
\newblock Underactuated robotics: Algorithms for walking, running, swimming,
  flying, and manipulation, 2023.

\bibitem{aerial_task:P&P}
Daniel Mellinger, Quentin Lindsey, Michael Shomin, and Vijay Kumar.
\newblock Design, modeling, estimation and control for aerial grasping and
  manipulation.
\newblock In {\em 2011 IEEE/RSJ International Conference on Intelligent Robots
  and Systems}, pages 2668--2673, 2011.

\bibitem{ODAR}
S.~{Park} et~al.
\newblock Odar: Aerial manipulation platform enabling omnidirectional wrench
  generation.
\newblock {\em IEEE/ASME Transactions on Mechatronics}, 23(4):1907--1918, Aug
  2018.

\bibitem{aerial_task:Slide}
Burak Y{\"u}ksel, Cristian Secchi, Heinrich~H B{\"u}lthoff, and Antonio
  Franchi.
\newblock Aerial physical interaction via ida-pbc.
\newblock {\em The International Journal of Robotics Research}, 38(4):403--421,
  2019.

\bibitem{aerial_task:PH}
Markus Ryll, Giuseppe Muscio, Francesco Pierri, Elisabetta Cataldi, Gianluca
  Antonelli, Fabrizio Caccavale, Davide Bicego, and Antonio Franchi.
\newblock 6d interaction control with aerial robots: The flying end-effector
  paradigm.
\newblock {\em The International Journal of Robotics Research},
  38(9):1045--1062, 2019.

\bibitem{aerial_task:Repair}
Pisak Chermprayong, Ketao Zhang, Feng Xiao, and Mirko Kovac.
\newblock An integrated delta manipulator for aerial repair: A new aerial
  robotic system.
\newblock {\em IEEE Robotics \& Automation Magazine}, 26(1):54--66, 2019.

\bibitem{perching:cup}
H.~W. Wopereis, T.~D. van~der Molen, T.~H. Post, S.~Stramigioli, and
  M.~Fumagalli.
\newblock Mechanism for perching on smooth surfaces using aerial impacts.
\newblock In {\em 2016 IEEE International Symposium on Safety, Security, and
  Rescue Robotics (SSRR)}, pages 154--159, 2016.

\bibitem{perching:cup2}
Sensen Liu, Wei Dong, Zhao Ma, and Xinjun Sheng.
\newblock Adaptive aerial grasping and perching with dual elasticity combined
  suction cup.
\newblock {\em IEEE Robotics and Automation Letters}, 5(3):4766--4773, 2020.

\bibitem{perching:spine}
Hai-Nguyen Nguyen, Robert Siddall, Brett Stephens, Alberto Navarro-Rubio, and
  Mirko Kovač.
\newblock A passively adaptive microspine grapple for robust, controllable
  perching.
\newblock In {\em 2019 2nd IEEE International Conference on Soft Robotics
  (RoboSoft)}, pages 80--87, 2019.

\bibitem{perching:spine2}
Hao Jiang, Shiquan Wang, and Mark~R Cutkosky.
\newblock Stochastic models of compliant spine arrays for rough surface
  grasping.
\newblock {\em The International Journal of Robotics Research}, 37(7):669--687,
  2018.

\bibitem{perching:adhesive}
Ludovic Daler, Adam Klaptocz, Adrien Briod, Metin Sitti, and Dario Floreano.
\newblock A perching mechanism for flying robots using a fibre-based adhesive.
\newblock In {\em 2013 IEEE International Conference on Robotics and
  Automation}, pages 4433--4438, 2013.

\bibitem{perching:dry-adhesive}
Justin Thomas, Morgan Pope, Giuseppe Loianno, Elliot~W Hawkes, Matthew~A
  Estrada, Hao Jiang, Mark~R Cutkosky, and Vijay Kumar.
\newblock Aggressive flight with quadrotors for perching on inclined surfaces.
\newblock {\em Journal of Mechanisms and Robotics}, 8(5):051007, 2016.

\bibitem{perching:gripper}
Haijie Zhang, Elisha Lerner, Bo~Cheng, and Jianguo Zhao.
\newblock Compliant bistable grippers enable passive perching for micro aerial
  vehicles.
\newblock {\em IEEE/ASME Transactions on Mechatronics}, 26(5):2316--2326, 2021.

\bibitem{perching:gripper2}
Tzu-Jui Lin, Siyu Long, and Karl~A. Stol.
\newblock Automated perching of a multirotor uav atop round timber posts.
\newblock In {\em 2018 IEEE/ASME International Conference on Advanced
  Intelligent Mechatronics (AIM)}, pages 486--491, 2018.

\bibitem{perching:bird-like-gripper}
William~RT Roderick, Mark~R Cutkosky, and David Lentink.
\newblock Bird-inspired dynamic grasping and perching in arboreal environments.
\newblock {\em Science Robotics}, 6(61):eabj7562, 2021.

\bibitem{perching:magnet}
Kazuaki Yanagimura, Kazunori Ohno, Yoshito Okada, Eijiro Takeuchi, and Satoshi
  Tadokoro.
\newblock Hovering of mav by using magnetic adhesion and winch mechanisms.
\newblock In {\em 2014 IEEE International Conference on Robotics and Automation
  (ICRA)}, pages 6250--6257, 2014.

\bibitem{perching:magnet2}
F.~J. Garcia-Rubiales, P.~Ramon-Soria, B.C. Arrue, and A.~Ollero.
\newblock Magnetic detaching system for modular uavs with perching capabilities
  in industrial environments.
\newblock In {\em 2019 Workshop on Research, Education and Development of
  Unmanned Aerial Systems (RED UAS)}, pages 172--176, 2019.

\bibitem{perching:rotor-suction}
Antonio~E. Jimenez-Cano, Pedro~J. Sanchez-Cuevas, Pedro Grau, Anibal Ollero,
  and Guillermo Heredia.
\newblock Contact-based bridge inspection multirotors: Design, modeling, and
  control considering the ceiling effect.
\newblock {\em IEEE Robotics and Automation Letters}, 4(4):3561--3568, 2019.

\bibitem{ceiling_effect_analysis}
Vernon~J Rossow.
\newblock Effect of ground and/or ceiling planes on thrust of rotors in hover.
\newblock 1985.

\bibitem{ceiling_effect_analysis2}
Xinkuang Wang, Shanshan Du, and Yong Liu.
\newblock Research on ceiling effect of quadrotor.
\newblock In {\em 2017 IEEE 7th Annual International Conference on CYBER
  Technology in Automation, Control, and Intelligent Systems (CYBER)}, pages
  846--851. IEEE, 2017.

\bibitem{ceiling_effect_analysis3}
Yasutada Tanabe, Masahiko Sugiura, Takashi Aoyama, Hideaki Sugawara, Shigeru
  Sunada, Koichi Yonezawa, and Hiroshi Tokutake.
\newblock Multiple rotors hovering near an upper or a side wall.
\newblock {\em Journal of Robotics and Mechatronics}, 30(3):344--353, 2018.

\bibitem{ceiling_effect_analysis4}
Stephen~A Conyers, Matthew~J Rutherford, and Kimon~P Valavanis.
\newblock An empirical evaluation of ceiling effect for small-scale rotorcraft.
\newblock In {\em 2018 International Conference on Unmanned Aircraft Systems
  (ICUAS)}, pages 243--249. IEEE, 2018.

\bibitem{ceiling_effect_model}
Yi~Hsuan Hsiao and Pakpong Chirarattananon.
\newblock Ceiling effects for surface locomotion of small rotorcraft.
\newblock In {\em 2018 IEEE/RSJ International Conference on Intelligent Robots
  and Systems (IROS)}, pages 6214--6219. IEEE, 2018.

\bibitem{ceiling_effect_model2}
Takuzumi Nishio, Moju Zhao, Fan Shi, Tomoki Anzai, Kento Kawaharazuka, Kei
  Okada, and Masayuki Inaba.
\newblock Stable control in climbing and descending flight under upper walls
  using ceiling effect model based on aerodynamics.
\newblock In {\em 2020 IEEE International Conference on Robotics and Automation
  (ICRA)}, pages 172--178. IEEE, 2020.

\bibitem{flight-array}
Raymond Oung, Frédéric Bourgault, Matthew Donovan, and Raffaello D'Andrea.
\newblock The distributed flight array.
\newblock In {\em 2010 IEEE International Conference on Robotics and
  Automation}, pages 601--607, 2010.

\bibitem{flying-gripper}
Bruno Gabrich, David Saldana, Vijay Kumar, and Mark Yim.
\newblock A flying gripper based on cuboid modular robots.
\newblock In {\em 2018 IEEE International Conference on Robotics and Automation
  (ICRA)}, pages 7024--7030. IEEE, 2018.

\bibitem{hydrus}
Moju Zhao, Koji Kawasaki, Tomoki Anzai, Xiangyu Chen, Shintaro Noda, Fan Shi,
  Kei Okada, and Masayuki Inaba.
\newblock Transformable multirotor with two-dimensional multilinks: Modeling,
  control, and whole-body aerial manipulation.
\newblock {\em The International Journal of Robotics Research},
  37(9):1085--1112, 2018.

\bibitem{jet-hr2}
Yuhang Li, Yuhao Zhou, Junbin Huang, Zijun Wang, Shunjie Zhu, Kairong Wu,
  Li~Zheng, Jiajin Luo, Rui Cao, Yun Zhang, and Zhifeng Huang.
\newblock Jet-hr2: A flying bipedal robot based on thrust vector control.
\newblock {\em IEEE Robotics and Automation Letters}, 7(2):4590--4597, 2022.

\bibitem{dragon}
Moju Zhao, Tomoki Anzai, Fan Shi, Xiangyu Chen, Kei Okada, and Masayuki Inaba.
\newblock Design, modeling, and control of an aerial robot dragon: A
  dual-rotor-embedded multilink robot with the ability of
  multi-degree-of-freedom aerial transformation.
\newblock {\em IEEE Robotics and Automation Letters}, 3(2):1176--1183, 2018.

\bibitem{lasdra}
Hyunsoo Yang, Sangyul Park, Jeongseob Lee, Joonmo Ahn, Dongwon Son, and Dongjun
  Lee.
\newblock Lasdra: Large-size aerial skeleton system with distributed rotor
  actuation.
\newblock In {\em 2018 IEEE International Conference on Robotics and Automation
  (ICRA)}, pages 7017--7023. IEEE, 2018.

\bibitem{zmp}
Miomir Vukobratovi{\'c} and J~Stepanenko.
\newblock On the stability of anthropomorphic systems.
\newblock {\em Mathematical biosciences}, 15(1-2):1--37, 1972.

\bibitem{jet-hr1}
Zhifeng Huang, Biao Liu, Jiapeng Wei, Qingsheng Lin, Jun Ota, and Yun Zhang.
\newblock Jet-hr1: Two-dimensional bipedal robot step over large obstacle based
  on a ducted-fan propulsion system.
\newblock In {\em 2017 IEEE-RAS 17th International Conference on Humanoid
  Robotics (Humanoids)}, pages 406--411, 2017.

\bibitem{leo}
Kyunam Kim, Patrick Spieler, Elena-Sorina Lupu, Alireza Ramezani, and Soon-Jo
  Chung.
\newblock A bipedal walking robot that can fly, slackline, and skateboard.
\newblock {\em Science Robotics}, 6(59):eabf8136, 2021.

\bibitem{omni}
Dario Brescianini and Raffaello D'Andrea.
\newblock Design, modeling and control of an omni-directional aerial vehicle.
\newblock In {\em 2016 IEEE international conference on robotics and automation
  (ICRA)}, pages 3261--3266. IEEE, 2016.

\bibitem{manipulability}
Tsuneo Yoshikawa.
\newblock Manipulability of robotic mechanisms.
\newblock {\em The international journal of Robotics Research}, 4(2):3--9,
  1985.

\bibitem{dynamic_manipulability}
Morteza Azad, Jan Babi{\v{c}}, and Michael Mistry.
\newblock Effects of the weighting matrix on dynamic manipulability of robots.
\newblock {\em Autonomous Robots}, 43(7):1867--1879, 2019.

\bibitem{velocity_damper}
Fumio Kanehiro, Mitsuharu Morisawa, Wael Suleiman, Kenji Kaneko, and Eiichi
  Yoshida.
\newblock Reactive leg motion generation method under consideration of physical
  constraints.
\newblock {\em Journal of the Robotics Society of Japan}, 28(10):1251--1261,
  2010.

\end{thebibliography}
